\def \beq {\begin{equation}}
\def \eeq {\end{equation}}
\def \beqa {\begin{eqnarray}}
\def \eeqa {\end{eqnarray}}
\def \lf {\left}
\def \ri {\right}
\def \sw {{$\sigma\!-\!\omega$} }
\newcommand{\dr}[1]{_{\rm #1}}
\newcommand{\req}[1]{(\ref{#1})}
\newcommand{\ANP}[1]{Adv.\ Nucl.\ Phys.\ #1}
\newcommand{\AP}[1]{Ann.\ Phys.\ (N.Y.) #1}
\newcommand{\ADNDT}[1]{At.\ Data Nucl.\ Data Tables #1}

\newcommand{\NPA}[1]{Nucl.\ Phys.\ A #1}
\newcommand{\PLB}[1]{Phys.\ Lett.\ B #1}

\newcommand{\PRC}[1]{Phys.\ Rev.\ C #1}
\newcommand{\PRB}[1]{Phys.\ Rev.\ B #1}
\newcommand{\PRA}[1]{Phys.\ Rev.\ A #1}

\newcommand{\PRp}[1]{Phys.\ Rep.\ #1}
\newcommand{\RPP}[1]{Rep.\ Prog.\ Phys.\ #1}
\newcommand{\ZPA}[1]{Z. Phys.\  A #1}

\newcommand{\IJMPE}[1]{Int.\ J. of Mod.\ Phys.\  E #1}
\documentstyle[12pt]{article}

\makeatletter
 \@addtoreset{equation}{section}
 
\makeatother
\makeatletter
\newcount\@tempcntc
\def\@citex[#1]#2{\if@filesw\immediate\write%
                  \@auxout{\string\citation{#2}}\fi
  \@tempcnta\z@\@tempcntb\m@ne\def\@citea{}\@cite{\@for\@citeb:=#2\do
    {\@ifundefined
       {b@\@citeb}{\@citeo\@tempcntb\m@ne\@citea%
                   \def\@citea{,}{\bf ?}\@warning
       {Citation `\@citeb' on page \thepage \space undefined}}%
    {\setbox\z@\hbox{\global\@tempcntc0\csname b@\@citeb%
                     \endcsname\relax}%
     \ifnum\@tempcntc=\z@ \@citeo\@tempcntb\m@ne
       \@citea\def\@citea{,}\hbox{\csname b@\@citeb\endcsname}%
     \else
      \advance\@tempcntb\@ne
      \ifnum\@tempcntb=\@tempcntc
      \else\advance\@tempcntb\m@ne\@citeo
      \@tempcnta\@tempcntc\@tempcntb\@tempcntc\fi\fi}}\@citeo}{#1}}
\def\@citeo{\ifnum\@tempcnta>\@tempcntb\else\@citea\def\@citea{,}%
  \ifnum\@tempcnta=\@tempcntb\the\@tempcnta\else
   {\advance\@tempcnta\@ne\ifnum\@tempcnta=\@tempcntb%
     \else \def\@citea{--}\fi
    \advance\@tempcnta\m@ne\the\@tempcnta\@citea\the\@tempcntb}\fi\fi}
\makeatother
\begin{document}
%
\vspace*{-1.0cm}
\begin{center}
{\Large\bf Nuclear surface properties in        \\[5mm]
           relativistic effective field theory}
\\[3.0cm]
M. Del Estal, M. Centelles, X. Vi\~nas  \\[2mm]
{\it Departament d'Estructura i Constituents de la Mat\`eria,
     Facultat de F\'{\i}sica,
\\
     Universitat de Barcelona,
     Diagonal {\sl 647}, E-{\sl 08028} Barcelona, Spain}
\end{center}
%
\vspace*{2.0cm}
\begin{abstract}
We perform Hartree calculations of symmetric and asymmetric
semi-infinite nuclear matter in the framework of relativistic models
based on effective hadronic field theories as recently proposed in the
literature. In addition to the conventional cubic and quartic scalar
self-interactions, the extended models incorporate a quartic vector
self-interaction, scalar-vector non-linearities and tensor couplings
of the vector mesons. We investigate the implications of these terms
on nuclear surface properties such as the surface energy coefficient,
surface thickness, surface stiffness coefficient, neutron skin
thickness and the spin--orbit force.
\end{abstract}

\mbox{}



\pagebreak
%
\section{Introduction}
\hspace*{\parindent}
Quantum hadrodynamics (QHD) and the relativistic treatment of nuclear
systems has been a subject of growing interest during recent years
\cite{Ser86,Cel86,Rei89,Ser92,Ser97}. The \sw model of Walecka
\cite{Ser86} and its non-linear extensions with cubic and quartic
self-interactions of the scalar-meson field \cite{Bog77} have been
widely used to this end. This model contains Dirac nucleons together
with neutral scalar and vector mesons as well as isovector-vector
$\rho$ mesons. At the mean field (Hartree) level, it already includes
the spin--orbit force, the finite range and the density dependence
which are essential ingredients of the nuclear interaction. This
simple model has become very popular in relativistic calculations and
describes successfully many properties of the atomic nucleus.

From a theoretical point of view, the non-linear \sw model with cubic
and quartic scalar self-interactions was classed within renormalizable
field theories which can be characterized by a finite number of
coupling constants. However, very recently, generalizations of this
model that include other non-linear interactions among the meson
fields and tensor couplings have been presented on the basis of
effective field theories by Serot et al.\
\cite{Ser97,Fur96,Mul96,Fur97}. The effective theory contains many
couplings of non-renormalizable form that are consistent with the
underlying symmetries of QCD. Consequently, one must find some
suitable expansion parameters and develop a systematic truncation
scheme. For this purpose the concept of naturalness has been employed:
it means that the unknown couplings of the theory should all be of the
order of unity when written in appropriate dimensionless form using
naive dimensional analysis \cite{Ser97,Fur96,Mul96,Fur97}. Then, one
can estimate the contributions coming from different terms by counting
powers in the expansion parameters and truncating the Lagrangian at a
given level of accuracy.

One important fact is the observation that at normal nuclear densities
the scalar and vector meson fields, denoted by $\Phi$ and $W$, are
small as compared with the nucleon mass $M$ and that they change
slowly in finite nuclei. This implies that the ratios $\Phi/M$, $W/M$,
$|\mbox{\boldmath $\nabla$}\Phi|/M^2$ and $|\mbox{\boldmath $\nabla$}
W|/M^2$ are useful expansion parameters when the effective field
theory is applied to the nuclear many-body problem. From this
viewpoint, if all the terms involving scalar and meson
self-interactions are retained in the Lagrangian up to fourth order,
one recovers the well-known non-linear \sw model plus some additional
terms \cite{Ser97,Fur96,Fur97}. For the truncation to be consistent,
the corresponding coupling constants should exhibit naturalness and
cannot be arbitrarily dropped out without an additional symmetry
argument. The effective Lagrangian truncated at fourth order contains
thirteen free parameters that have been fitted to reproduce
twenty-nine finite nuclei observables \cite{Ser97,Fur97}. Remarkably,
the fitted parameters turn out to be natural and the results are not
dominated by the last terms retained. This evidence confirms the
utility of the principles of naive dimensional analysis and
naturalness and shows that truncating the effective Lagrangian at the
first lower orders is justified.

The term with a vector-meson quartic self-interaction has been
considered previously in relativistic mean field (RMF) calculations
from a phenomenological point of view. Bodmer \cite{Bod91} considered
this coupling to avoid the negative coefficient of the quartic scalar
self-interaction that appears in many non-linear \sw parametrizations
that correctly describe the atomic nucleus \cite{Bod89}. In some
special situations this negative term can lead to a pathological
behaviour of the scalar potential. On the other hand, the equation of
state is softened at moderate high densities when the vector
non-linearity is taken into account. The quartic vector
self-interaction has also been phenomenologically used by Gmuca
\cite{Gmu92a,Gmu92b} in a non-linear \sw model for parametrizing
Dirac--Brueckner--Hartree--Fock calculations of nuclear matter. The
same idea was developed by Toki et al.\ and applied to study finite
nuclei \cite{Sug94} and neutron stars \cite{Sum95}. Recently, the
properties of high-density nuclear and neutron matter have been
analyzed in the RMF approach taking into account scalar and vector
non-linearities \cite{Mul96}.

The tensor couplings of the vector $\omega$ and $\rho$ mesons to the
nucleon were investigated by Reinhard et al.\ \cite{Rei89,Ruf88} as an
extension of the RMF model, and more recently by Furnstahl et al.\
\cite{Fur97,Fur98} from the point of view of relativistic effective
field theory. In these works it was shown that the tensor coupling of
the $\omega$ meson has an important bearing on the nuclear spin--orbit
splitting.

The surface properties of nuclei play a crucial role in certain
situations. This is the case, for instance, of saddle-point
configurations in nuclear fission or fragment distributions in
heavy-ion collisions. Within a context related to the liquid droplet
model (LDM) and the leptodermous expansion \cite{Mye69}, the surface
properties can be extracted from semi-infinite nuclear matter
calculations either quantally or semiclassically (though the total
curvature energy coefficient can only be computed semiclassically
\cite{Cen96}). In the non-relativistic case most of the calculations
of the surface properties have been carried out using Skyrme forces,
quantally \cite{Bra85} or semiclassically with the help of the
extended Thomas--Fermi (ETF) method \cite{Bra85,Sto88}. In the
relativistic case the nuclear surface has been analyzed within the \sw
model since a long time ago. The calculations have been performed
semiclassically using the relativistic Thomas--Fermi (TF) method or
its extensions (RETF), for symmetric
\cite{Ser86,Bog77,Sto91,Cen93,Spe93} and asymmetric
\cite{Von94a,Cen98} matter, and also in the quantal Hartree approach
\cite{Hof89,Von94b,Von95}.

In the framework of the relativistic model and effective field theory,
the main purpose of the present work is to carefully analyze the
influence on surface properties of the quartic vector non-linearity,
of the newly proposed scalar-vector self-interactions and of the
tensor coupling. We shall investigate quantities such as the surface
energy coefficient, surface thickness, spin--orbit strength, surface
stiffness coefficient and neutron skin thickness obtained from Hartree
calculations of symmetric and asymmetric semi-infinite nuclear matter.

The paper is organized as follows. Section 2 is devoted to the basic
theory. The results on the surface properties of symmetric matter are
discussed in Section 3. Section 4 addresses the case of asymmetric
systems. The summary and conclusions are given in the last section.

\pagebreak
%
\section{Mean field equations for symmetric semi-infinite nuclear
         matter}
\hspace*{\parindent}
Following Ref.\ \cite{Fur96}, to derive the mean field equations one
starts from an energy functional containing Dirac baryons and
classical scalar and vector mesons. The energy functional can be
obtained from the effective Lagrangian in the Hartree approach using
many-body techniques \cite{Ser97,Fur97}. However, this energy
functional can also be considered as an expansion in $\Phi/M$, $W/M$,
$|\mbox{\boldmath $\nabla$}\Phi|/M^2$ and $|\mbox{\boldmath $\nabla$}
W|/M^2$ of a general energy density functional that contains all the
correlation effects. The theoretical basis of this functional lies on
the extension of the Hohenberg--Kohn theorem \cite{Hoh64} to QHD
\cite{Spe92}. Using the Kohn--Sham scheme \cite{Koh65} with the mean
fields playing the role of Kohn--Sham potentials, one finds similar
mean field equations to those obtained from the Lagrangian
\cite{Spe92}, but including effects beyond the Hartree approach
through the non-linear couplings \cite{Ser92,Fur96,Mul96,Fur97}.

A semi-infinite system of uncharged nucleons corresponds to a
one-dimensional geometry where half the space is filled with nuclear
matter at saturation and the other half is empty, so that a surface
develops around the interface. The fields and densities change only
along the direction perpendicular to the medium. Specifying the energy
density functional considered in Refs.\ \cite{Ser97} and \cite{Fur97}
to symmetric semi-infinite nuclear matter with the surface normal
pointing into the $z$ direction one has
%
\beqa
{\cal E}(z) & = &  \sum_\alpha \varphi_\alpha^\dagger(z)
\lf\{ -i \mbox{\boldmath$\alpha$} \!\cdot\! \mbox{\boldmath$\nabla$}
+ \beta [M - \Phi(z)] + W(z)
- \frac{i f\dr{v}}{2M} \, \beta
  \mbox{\boldmath$\alpha$} \!\cdot\! \mbox{\boldmath$\nabla$} W(z)
\ri\} \varphi_\alpha (z)
\nonumber \\[3mm]
& & \null + \frac{1}{2g\dr{s}^2}\lf( 1 +
\alpha_1\frac{\Phi(z)}{M}\ri) \lf(
\mbox{\boldmath $\nabla$}\Phi(z)\ri)^2
+ \lf ( \frac{1}{2}
+ \frac{\kappa_3}{3!}\frac{\Phi(z)}{M}
+ \frac{\kappa_4}{4!}\frac{\Phi^2(z)}{M^2}\ri )
 \frac{m\dr{s}^2}{g\dr{s}^2} \Phi^2(z)
\nonumber \\[3mm]
& & \null - \frac{1}{2g\dr{v}^2}\lf( 1 +\alpha_2\frac{\Phi(z)}{M}\ri)
\lf( \mbox{\boldmath $\nabla$} W(z)  \ri)^2 -
\frac{\zeta_0}{4!} \frac{1}{ g\dr{v}^2 } W^4 (z)
\nonumber \\[3mm]
& &  \null - \frac{1}{2}\lf(1 + \eta_1 \frac{\Phi(z)}{M} +
\frac{\eta_2}{2} \frac{\Phi^2 (z)}{M^2} \ri)
 \frac{m\dr{v}^2}{g\dr{v}^2} W^2 (z) \,,
\label{eq1}
\eeqa
where the index $\alpha$ runs over all occupied states of the positive
energy spectrum, $\Phi \equiv g\dr{s} \phi_0$ and $ W \equiv g\dr{v}
V_0$ (notation as in Ref.\ \cite{Ser86}). Except for the terms with
$\alpha_1$ and $\alpha_2$, the functional \req{eq1} is of fourth order
in the expansion. We retain the fifth-order terms $\alpha_1$ and
$\alpha_2$ because in Refs.\ \cite{Ser97} and \cite{Fur97} they have
been estimated to be numerically of the same magnitude as the quartic
scalar term in the nuclear surface energy.

The mean field equations are obtained by minimizing with respect to
$\varphi^\dagger_\alpha$, $\Phi$ and $W$:
%
\beq
\lf\{ -i \mbox{\boldmath$\alpha$} \!\cdot\! \mbox{\boldmath$\nabla$}
 + \beta [M - \Phi(z)] + W(z)
 - \frac{i f\dr{v}}{2M} \, \beta
  \mbox{\boldmath$\alpha$} \!\cdot\! \mbox{\boldmath$\nabla$} W(z)
 \ri\} \varphi_\alpha (z) =
\varepsilon_\alpha \, \varphi_\alpha (z) \,,
\label{eq2}
\eeq
%
\beqa
   -\Delta \Phi(z) + m\dr{s}^2 \Phi(z)  & = &
     g\dr{s}^2 \rho\dr{s}(z)
        -{m\dr{s}^2\over M}\Phi^2 (z)
         \lf({\kappa_3\over 2}+{\kappa_4\over 3!}{\Phi(z)\over M}
               \ri )
                \nonumber  \\[3mm]
     & & \null
       +{g\dr{s}^2 \over 2M}
         \lf(\eta_1+\eta_2{\Phi(z)\over M}\ri)
                { m\dr{v}^2\over  g\dr{v}^2} W^2 (z)
          \nonumber  \\[3mm]
     & & \null
       +{\alpha_1 \over 2M}[
             (\mbox{\boldmath $\nabla$}\Phi(z))^2
             +2\Phi(z)\Delta \Phi(z) ]
        + {\alpha_2 \over 2M} {g\dr{s}^2\over g\dr{v}^2}
        (\mbox{\boldmath $\nabla$}W(z))^2 ,
 \label{eq3}  \\[3mm]
 -\Delta W(z) +  m\dr{v}^2 W(z)  & = &
 g\dr{v}^2 \lf( \rho(z) + \frac{f\dr{v}}{2} \rho\dr{T}(z) \ri)
 -\lf( \eta_1+{\eta_2\over 2}{\Phi(z)\over M} \ri ){\Phi(z)\over M}
 m\dr{v}^2 W(z)        \nonumber  \\[3mm]
         & & \null
      -{1\over 3!}\zeta_0 W^3(z)
          +{\alpha_2 \over M} [\mbox{\boldmath $\nabla$}\Phi(z)
\cdot\mbox{\boldmath $\nabla$}W(z)
                    +\Phi(z)\Delta W(z)] \,.
\label{eq4}
\eeqa
The baryon, scalar and tensor densities are respectively
%
\beqa
 \rho(z) & = &
 \sum_\alpha \varphi_\alpha^\dagger(z) \varphi_\alpha(z) \,,
\label{eq4a} \\[3mm]
 \rho\dr{s}(z) & = &
 \sum_\alpha \varphi_\alpha^\dagger(z) \beta \varphi_\alpha(z) \,,
\label{eq4b} \\[3mm]
 \rho\dr{T}(z) & = &
 \sum_\alpha \frac{i}{M} \mbox{\boldmath$\nabla$} \!\cdot\!
 \lf[ \varphi_\alpha^\dagger(z) \beta \mbox{\boldmath$\alpha$}
      \varphi_\alpha(z) \ri] .
\label{eq4c}
\eeqa
The expression of the four-component spinors $\varphi_\alpha(z)$ in
the semi-infinite medium was given by Hofer and Stocker in Ref.\
\cite{Hof89}.

In a semi-infinite nuclear matter calculation the sum over the
single-particle states is replaced by an integration over momenta:
 \beq
 \sum_\alpha \longrightarrow 2 \frac{\Omega}{(2\pi)^3} \sum_{\lambda}
 \int d{\bf{k}}  \,,
 \label{eq5}
\eeq
where $\Omega$ stands for the volume of the box, the factor 2 takes
into account the isospin degree of freedom and $\lambda$ describes the
spin orientation of the nucleons. Introducing the Fermi momentum
$k\dr{F}$, the integration domain is restricted to $k_{x}^2 + k_{y}^2
+ k_{z}^2 = k_{\bot}^2 + k_{z}^2 \leq k\dr{F}^2$, with $k_{z} \geq 0$
if the bulk nuclear matter is located at $z=-\infty$. Following the
method outlined in Ref.\ \cite{Hof89} one finds two sets ($\lambda =
\pm1$) of first-order differential equations for the orbital part of
the upper and lower components of the Dirac spinors:
%
\beqa
\frac{dG_a (z)}{dz}
- \lf[ \lambda k_\bot + \frac{f\dr{v}}{2M} \frac{dW(z)}{dz} \ri]
  G_a (z) & = &
\lf[\varepsilon_a - W(z) + M^*(z)\ri] F_a (z) \,,
\label{eq6a} \\[3mm]
- \frac{dF_a (z)}{dz}
- \lf[ \lambda k_\bot + \frac{f\dr{v}}{2M} \frac{dW(z)}{dz} \ri]
  F_a (z) & = &
\lf[\varepsilon_a - W(z) - M^*(z)\ri] G_a (z) \,,
\label{eq6b}\eeqa
where $a = (k_{z},k\dr{\bot},\lambda)$ and $M^{*}(z) = M - \Phi(z)$ is
the Dirac effective mass of the nucleons. From the asymptotic
behaviour at $z=-\infty$ (bulk nuclear matter), the condition on the
energy eigenvalues is $\varepsilon_a = \sqrt{k_{\bot}^2 + k_{z}^2 +
{M^{*}_{\infty}}^{2}} + W_{\infty}$, with $M^{*}_{\infty}$ and
$W_{\infty}$ being the nuclear matter values of $M^{*}$ and $W$. The
densities for each spin orientation $\lambda= \pm1$ read
%
\beqa
\rho^\lambda (z) & = & \frac{2}{\pi^2} \int_{0}^{ k
\dr{F}} dk_{z} \int_{0}^{\sqrt{k\dr{F}^2 - k_z^2}}
 dk_{\bot} k_{\bot} \lf( \lf |G_a (z)
\ri |^2 + \lf |F_a (z) \ri |^2 \ri )  ,
\label{eq7}
\\[3mm]
\rho\dr{s}^\lambda (z) & = & \frac{2}{\pi^2} \int_{0}^{ k
\dr{F}} dk_{z} \int_{0}^{\sqrt{k\dr{F}^2 - k_z^2}}
 dk_{\bot} k_{\bot} \lf( \lf |G_a (z)
\ri |^2 - \lf |F_a (z) \ri |^2 \ri ) ,
\label{eq8}
\\[3mm]
\rho\dr{T}^\lambda (z) & = & \frac{2}{\pi^2} \int_{0}^{ k
\dr{F}} dk_{z} \int_{0}^{\sqrt{k\dr{F}^2 - k_z^2}} dk_{\bot} k_{\bot}
 \frac{d}{dz} \lf( \frac{2}{M} F_a(z) G_a(z) \ri) ,
\label{eq9} \eeqa
and the total densities are given by
%
\beq
\rho(z) = \sum_\lambda \rho^\lambda (z) \,, \qquad
\rho\dr{s}(z) = \sum_\lambda \rho\dr{s}^\lambda (z) \,, \qquad
\rho\dr{T}(z) = \sum_\lambda \rho\dr{T}^\lambda (z) \,.
\label{eq9b}\eeq

Using the equations of motion the energy density of the semi-infinite
nuclear matter system can be written as follows:
%
\beqa
{\cal E}(z) & = & \frac{2}{\pi^2} \sum_{\lambda} \int_{0}^{k
\dr{F}} dk_{z} \int_{0}^{\sqrt{k\dr{F}^2 - k_z^2}}
 dk_{\bot} k_{\bot}
 \lf( \sqrt{k_{\bot}^2 + k_{z}^2 + {M^{*}_{\infty}}^2}+W_{\infty} \ri)
 \lf (\lf | G_a (z) \ri |^2 + \lf |F_a (z) \ri |^2 \ri )
\nonumber \\[3mm]
& & \null + \frac{1}{2} \Phi(z)\rho\dr{s}(z)
- \frac{1}{2} W(z) \lf( \rho(z) + \frac{f\dr{v}}{2} \rho\dr{T}(z) \ri)
- \frac{\Phi(z)}{4M}
\lf(\frac{\kappa_3}{3}+\frac{\kappa_4}{6} \frac{\Phi(z)}{M} \ri)
\frac{m\dr{s}^2}{ g\dr{s}^2} \Phi^2 (z)
 \nonumber   \\[3mm]
& & \null + \frac{\Phi(z)}{4M}
\lf( \eta_1 +\eta_2 \frac{\Phi(z)}{M} \ri)
\frac{ m\dr{v}^2}{g\dr{v}^2}W^2 (z) + \frac{\zeta_0}{4!}
\frac{1}{g\dr{v}^2}W^4 (z)
\nonumber \\[3mm]
& & \null
-\frac{\alpha_1}{4g\dr{s}^2} \frac{\Phi(z)}{M}
 \lf(\mbox{\boldmath $\nabla$} \Phi(z) \ri)^2
+ \frac{\alpha_2}{4g\dr{v}^2 }\frac{\Phi(z)}{M}
\lf( \mbox{\boldmath $\nabla$} W(z) \ri)^2 .
\label{eq10} \eeqa
Finally, the surface energy coefficient $E\dr{s}$ is obtained from the
expression \cite{Mye69}
\beq
 E\dr{s} = 4\pi r_{0}^2 \int^{\infty}_{-\infty} dz
 \lf[ {\cal E}(z) - (a\dr{v}+M) \rho(z) \ri] ,
\label{eq11} \eeq
where $a\dr{v}$ is the energy per particle in bulk nuclear matter and
$r_0$ is the nuclear radius constant: $r_0 = (3/4 \pi \rho_0)^{1/3}$,
with $\rho_0$ the nuclear matter density.

Another important quantity in the study of the nuclear surface
structure is the spin--orbit interaction. By elimination of the lower
spinor in terms of the upper spinor, one obtains a Schr\"odinger-type
equation with a term $V\dr{so}(z)$ that has the structure of the
single-particle spin--orbit potential for the non-relativistic case
\cite{Hof89,Von95}. In our present model the orbital part of the
spin--orbit potential reads as
%
\beq
 V\dr{so}(z) = \frac{1}{2M}
 \lf[ \frac{1}{\varepsilon_a - W(z) + M^*(z)}
 \lf( \frac{dW(z)}{dz} + \frac{d\Phi(z)}{dz} \ri)
 + \frac{f\dr{v}}{M} \frac{dW(z)}{dz}  \ri] .
\label{eq12}\eeq
In the non-relativistic limit, by means of a Foldy--Wouthuysen
reduction, Eq.\ \req{eq12} becomes
%
\beq
 V\dr{so}^{\rm FW}(z) = \frac{1}{4M^2}
 \lf[ (1 + 2 f\dr{v}) \frac{dW(z)}{dz} + \frac{d\Phi(z)}{dz} \ri]
\label{eq13}\eeq
and the nucleons are then moving in a central potential of the form
%
\beq
 V\dr{c}(z) = W(z) - \Phi(z) \,.
\label{eq14}\eeq

\pagebreak
%
\section{Surface properties in the symmetric case}
\hspace*{\parindent}
Although previous works
\cite{Cen93,Spe93,Von94a,Cen98,Hof89,Von94b,Von95} have thoroughly
investigated the properties of the nuclear surface in the standard
non-linear \sw model, for which $\kappa_3$ and $\kappa_4$ are the only
non-linearities in \req{eq1}, here we wish to enlarge this study by
including the additional non-linear and tensor couplings considered in
Refs.\ \cite{Ser97,Fur96,Fur97}. In concrete, we want to study the
role of the quartic vector self-interaction $\zeta_0$ that has been
used after the work of Bodmer \cite{Bod91}, the role of the terms with
$\eta_1$ and $\eta_2$ that couple the scalar and vector fields, of the
terms with $\alpha_1$ and $\alpha_2$ that imply the gradients of the
fields, and of the tensor coupling $f\dr{v}$ of the vector $\omega$
meson to the nucleon. While $\zeta_0$, $\eta_1$ and $\eta_2$ can be
classified as volume contributions, the couplings $\alpha_1$,
$\alpha_2$ and $f\dr{v}$ are genuine surface terms. On the basis of
the concept of naturalness there is no reason to omit any of these
terms in the energy density functional \req{eq1}, unless there exists
a symmetry principle to forbid it. However, to clarify the impact on
the surface properties of the aforementioned couplings we will analyze
each one separately, as it has been similarly done in Ref.\
\cite{Fur96}. In this section we shall study symmetric systems, while
in Section 4 we shall address the case of asymmetric matter.

%
%
\subsection{Effect of the quartic vector self-interaction}
\hspace*{\parindent}
In the conventional non-linear \sw model the value of the coefficients
$g\dr{s}^2/m\dr{s}^2$, $g\dr{v}^2/m\dr{v}^2$, $\kappa_3$ and
$\kappa_4$ can be univocally obtained by imposing that for nuclear
matter at saturation the density $\rho_0$, energy per particle
$a\dr{v}$, effective mass $M^{*}_{\infty}/M$ and incompressibility
modulus $K$ take given values. When the vector-meson quartic
self-interaction is switched on, the Dirac equation for the baryons
and the Klein--Gordon equation for the vector field in infinite
nuclear matter become (throughout this subsection we set
$\eta_1=\eta_2=\alpha_1=\alpha_2=f\dr{v}=0$):
%
\beqa
 a\dr{v} & = &
 \sqrt{k\dr{F}^2 + {M^{*}_{\infty}}^2} + W_{\infty} - M \,,
\label{eq3.1}  \\[3mm]
 m\dr{v}^2 W_{\infty} & = & g\dr{v}^2 \rho_0
 - \frac{1}{6}\zeta_0 W^3_{\infty} \,.
\label{eq3.2} \eeqa
The saturation density $\rho_0$ and the Fermi momentum $k\dr{F}$ are
related as usual by $\rho_0= 2k\dr{F}^3/3\pi^2$. Specifying $\rho_0$,
$a\dr{v}$ and $M^{*}_{\infty}/M$, from the above equations one
extracts the coupling constant $g\dr{v}$ as a function of $\zeta_0$
(the nucleon and $\omega$ masses take their empirical values: $M= 939$
MeV and $m\dr{v}= 783$ MeV). The steps to calculate
$g\dr{s}^2/m\dr{s}^2$, $\kappa_3$ and $\kappa_4$ are then the same as
when $\zeta_0= 0$, see e.g.\ Refs.\ \cite{Bod91} and \cite{Bod89}, but
now these coefficients become functions of $\zeta_0$. The reader will
find a detailed study of the implications of the vector non-linearity
$\zeta_0$ in nuclear matter in Refs.\ \cite{Fur96} and \cite{Bod91}.

The assumption of naturalness requires that the couplings
$g\dr{s}/4\pi$, $g\dr{v}/4\pi$, $\kappa_3$, $\kappa_4$ and $\zeta_0$
should all be roughly of the order of unity. Figure 1 illustrates the
variation of these couplings as a function of the non-dimensional
parameter
%
\beq
\eta_0 = \frac{m\dr{v}^2}{g\dr{v}^2}
\sqrt{ \frac{6 m\dr{v}^2}{\zeta_0 \rho_0^2} }
\label{eq3.2.b} \eeq
used by
Bodmer\footnote{Notice that in Ref.\ \cite{Bod91} the
parameter $\eta_0$ was called $z$.}
in Ref.\ \cite{Bod91}. With $m\dr{s}= 490$ MeV, the figure presents
the results for four sets of nuclear matter properties: $\rho_0 =
0.152$ fm$^{-3}$ ($k\dr{F}= 1.31$ fm$^{-1}$), $a\dr{v} = -16.42$ MeV,
$K= 200$ and 350 MeV, and $M_{\infty}^{*}/M= 0.6$ and 0.7. When $K=
200$ MeV and $M_{\infty}^{*}/M= 0.6$ the equilibrium properties of the
interaction and the scalar mass $m\dr{s}$ are very close to those of
the non-linear parametrization NL1 \cite{Rei86}.

We realize that $g\dr{s}$ and $g\dr{v}$ are only weakly affected by
$\eta_0$ (the change is not appreciable in the scale of Figure 1).
However, $\eta_0$ has a direct effect on the scalar-meson quartic
self-interaction $\kappa_4$, moving it from a negative value when the
quartic vector term is absent ($\eta_0 =\infty$) to a desirable
positive value when $\eta_0 \sim 2$. The change of sign of $\kappa_4$
takes place at larger values of $\eta_0$ for larger $K$ and
$M^{*}_{\infty}/M$; in fact, $\kappa_4$ is already positive at
$\eta_0= \infty$ if $K = 350$ MeV and $M^{*}_{\infty}/M = 0.7$. Both
non-linearities of the scalar field $\kappa_3$ and $\kappa_4$ remain
in the natural zone for $2 \leq \eta_0 \leq \infty$ approximately,
excepting the case of $K = 350$ MeV and $M^{*}_{\infty}/M = 0.7$, but
they start to depart appreciably from their natural values when
$\eta_0 \leq 2$. These trends fairly agree with the assumption of
naturalness: $\eta_0 \geq 2$ corresponds to the region where $\zeta_0$
can be considered as natural (see the figure), and for this range of
$\eta_0$ values also the rest of coupling constants are natural. The
behaviour gleaned from Figure 1 is rather independent of the
saturation density $\rho_0$ and energy $a\dr{v}$, and e.g.\ we have
found similar trends with the specific nuclear matter properties used
by Bodmer in Ref.\ \cite{Bod91}.

Next we turn our attention to the surface properties. In Figures 2 and
3 we have plotted, respectively, the surface energy coefficient
$E\dr{s}$ and the surface thickness $t$ of the semi-infinite density
profile (the 90\%--10\% fall-off distance) against the $\eta_0$
parameter. The selected values of $\eta_0$ are those employed in Table
1 of Ref.\ \cite{Bod91}. With $\rho_0 = 0.152$ fm$^{-3}$ and $a\dr{v}
= -16.42$ MeV, we have performed the calculations for a few
incompressibilities ($K= 125$, 200 and 350 MeV) and effective masses
($M^{*}_{\infty}/M= 0.6$ and 0.7). Furthermore, we have considered two
values of the mass of the scalar meson ($m\dr{s}= 490$ and 525 MeV).
This quantity governs the range of the attractive interaction and
determines the surface fall-off: a larger $m\dr{s}$ results in a
steeper surface and a reduction of $E\dr{s}$ and $t$.

From Figures 2 and 3 we realize that the quartic vector
self-interaction scarcely alters the values of the surface energy
coefficient and of the surface thickness if $\zeta_0$ remains in the
natural domain, i.e., if $2 \leq \eta_0 \leq \infty$. At $K= 350$ MeV
and $M^{*}/M \geq 0.6$, $E\dr{s}$ is raised by decreasing $\eta_0$.
This tendency may be inverted at smaller values of the
incompressibility, depending also on the value of the nucleon
effective mass, but the global trends are practically independent of
the mass of the scalar meson. The thickness $t$ exhibits a more
monotonous behaviour: in all cases it stays almost equal to its value
at $\eta_0= \infty$ and goes down slightly for $\eta_0 \leq 2$.
Noticeable departures of the surface energy coefficient and thickness
from the $\eta_0 = \infty$ values can be found only if $\eta_0$ is
decreased beyond the natural limit, a situation where the interaction
is mainly ruled by the vector-meson quartic self-interaction.

Even though the impact of the vector-meson quartic self-interaction is
small, it can help to find parameter sets for which both the surface
energy coefficient and the surface thickness lie in the empirical
region. This fact is illustrated in Figure 4, where $E\dr{s}$ and $t$
are drawn versus $M^{*}_{\infty}/M$ for $\eta_0= \infty$, 2 and $0.5$
and for $m\dr{s}= 450$, 500 and 550 MeV, using the nuclear matter
conditions of Ref.\ \cite{Bod91}: $\rho_0 = 0.1484$ fm$^{-3}$,
$a\dr{v} = -15.75$ MeV and $K=200$ MeV\@. (We also performed the
calculations for $\eta_0= 5$ and $\eta_0= 1$ to confirm the trends we
discuss below.) The horizontal dashed lines in Figure 4 serve to
indicate the empirical region for the surface energy and thickness.

If at $\eta_0= \infty$ we concentrate, for instance, on the
parametrizations with $m\dr{s}=450$ MeV we see that the one with
$M^{*}_{\infty}/M= 0.7$ yields, simultaneously, $E\dr{s}$ and $t$
within the empirical region. The agreement between the calculated
values with $m\dr{s}=450$ MeV and the empirical region improves for
$\eta_0= 2$, where both $E\dr{s}$ and $t$ are acceptable for
practically all the values of the effective mass considered.

For $\eta_0 \geq 2$ the dependence of $E\dr{s}$ and $t$ upon the
nucleon effective mass is similar to that found in the usual \sw model
without a quartic vector self-interaction \cite{Cen93,Von94b}. The
general tendencies start to change when $\eta_0$ is lowered and leaves
the natural region. As $\eta_0$ becomes smaller Figure 4 shows that
the slope of the curves of $E\dr{s}$ and $t$ as a function of
$M^{*}_{\infty}/M$ changes, and that the curves for the different
$m\dr{s}$ come closer together.

To get more insight about the influence of the vector-meson quartic
self-interaction, we display in Figure 5 the profiles of the baryon
density $\rho(z)$ and of the surface tension density (Swiatecki
integrand) $\sigma(z)= {\cal E}(z) - (a\dr{v}+M) \rho(z)$, Eq.\
\req{eq11}. In turn, we have represented in Figure 6 the orbital part
of the spin--orbit potential $V\dr{so}(z)$, Eq.\ \req{eq12}, at the
Fermi surface (i.e., evaluated at $k= k\dr{F}$) and the central mean
field $V\dr{c}(z)$ defined in Eq.\ \req{eq14}. The properties of the
interactions used in these figures are $\rho_0= 0.152$ fm$^{-3}$,
$a\dr{v}= -16.42$ MeV, $K= 200$ MeV, $M^{*}_{\infty}/M= 0.6$ and 0.7,
and $m\dr{s}= 490$ MeV\@. Results are shown for $\eta_0= \infty$, 2
and 0.5. The corresponding values of the surface energy coefficient
and surface thickness can be read from Figures 2 and 3.

It can be seen that the local quantities depicted in Figures 5 and 6
oscillate as functions of $z$ (Friedel oscillations). Both the surface
tension density and the single-particle spin--orbit potential are
confined to the surface and average to zero inwards as the bulk matter
is approached. The local profiles for $\eta_0= 2$, which somehow marks
the limit of naturalness as we have commented, are almost equal to
those obtained in the absence of the quartic vector self-interaction.
In agreement with Figures 2 and 3, only when $\eta_0$ is decreased to
non-natural values one can notice some changes in the profiles, which
are more visible for $M^{*}_{\infty}/M = 0.6$ than for
$M^{*}_{\infty}/M = 0.7$. Decreasing $\eta_0$ makes the surface
steeper and the thickness $t$ smaller, produces an enhancement in the
surface region of the density $\rho(z)$ and of the mean field
$V\dr{c}(z)$, and builds up Friedel oscillations in $\sigma(z)$ and in
the spin--orbit potential $V\dr{so}(z)$.

It is well known that the experimental spin--orbit splittings require
within narrow bounds a Dirac effective mass $M^{*}_{\infty}/M$ around
0.6 in the conventional relativistic model \cite{Bod89,Fur96}. Figure
6 shows that introducing a quartic vector self-interaction makes the
spin--orbit well deeper. However, it is only a minor effect: with
$M^{*}_{\infty}/M= 0.7$ it is not possible to reproduce the
spin--orbit interaction of the case $M^{*}_{\infty}/M= 0.6$ at
$\eta_0= \infty$, not even if one sets $\eta_0= 0.5$ (which in
addition brings about an unreallistically small $t$, see Figure 3). A
similar conclusion was drawn in Ref.\ \cite{Bod91} from an analysis in
nuclear matter. We also have computed the non-relativistic limit
$V\dr{so}^{\rm FW}(z)$ of the spin--orbit potential given by the
expression \req{eq13}. In agreement with Ref.\ \cite{Von95} we have
found that while $V\dr{so}^{\rm FW}(z)$ qualitatively reproduces the
behaviour of $V\dr{so}(z)$, it strongly underestimates the
quantitative depth of the fully relativistic spin--orbit strength (by
$\sim 20$\% for $M^{*}_{\infty}/M= 0.6$ and by $\sim 15$\% for
$M^{*}_{\infty}/M= 0.7$).

%
%
\subsection{Influence of the volume cubic and quartic scalar-vector
interactions}
\hspace*{\parindent}
The next bulk terms in the energy density \req{eq1} that contain
non-linear meson interactions are
%
\beq
- \frac{1}{2}\frac{\Phi}{M}
\lf( \eta_1 +\frac{\eta_2}{2} \frac{\Phi}{M} \ri)
\frac{m\dr{v}^2}{g\dr{v}^2} W^2  \,.
\label{eq3.7} \eeq
In analogy to Figure 1 for $\eta_0$, in Figure 7 we study the change
of the couplings $g\dr{s}/4\pi$, $g\dr{v}/4\pi$, $\kappa_3$ and
$\kappa_4$ with the parameters $\eta_1$ and $\eta_2$ (introducing each
one separately) for some specific equilibrium properties. With $\rho_0
= 0.152$ fm$^{-3}$, $a\dr{v} = -16.42$ and $m\dr{s}= 490$ MeV, in part
(a) of Figure 7 it is $K= 200$ MeV and $M_{\infty}^{*}/M= 0.6$, while
in part (b) it is $K= 350$ MeV and $M_{\infty}^{*}/M= 0.7$.

For $K= 200$ MeV and $M_{\infty}^{*}/M= 0.6$ both $\eta_1$ and
$\eta_2$ have a considerable effect on the scalar non-linearities
$\kappa_3$ and $\kappa_4$. The coupling $\kappa_4$ changes sign for
$\eta_2 \approx 1$, but it remains negative in the interval of
$\eta_1$ values used. To keep all the coupling constants within
natural values we see that the range for $\eta_1$ and $\eta_2$ (when
introduced separately) is restricted to run roughly from $-0.5$ to
$2.5$. Consider now different values of the incompressibility and
effective mass, as in part (b) of Figure 7. The dependence of the
couplings on $\eta_2$ is not significantly altered. However,
increasing either $K$ or $M_{\infty}^{*}/M$ results in a smoother
slope of $\kappa_3$ with $\eta_1$, while it makes $\kappa_4$ grow
steadily with $\eta_1$ and become positive at some value of this
parameter. We have checked for $K= 200$ MeV and $M_{\infty}^{*}/M=
0.55$, and for $K= 125$ MeV and $M_{\infty}^{*}/M= 0.6$, that
$\kappa_4$ remains negative at all values of $\eta_1$ and that
$\kappa_3$ grows with $\eta_1$ faster than in part (a) of Figure 7.

The influence of the non-linear interactions $\eta_1$ and $\eta_2$ on
the surface properties is analyzed in Figure 8. With $\zeta_0=
\alpha_1= \alpha_2= f\dr{v}= 0$, in this figure we have computed
$E\dr{s}$ and $t$ taking into account the terms \req{eq3.7}. The
saturation conditions of the interaction and the scalar mass are the
same as in part (a) of Figure 7. The results are displayed in the
plane $\eta_1$--$\eta_2$ in the form of contour plots of constant
$E\dr{s}$ (solid lines) and of constant $t$ (dashed lines). The range
of variation of $\eta_1$ and $\eta_2$ lies in the region imposed by
naturalness, and yields values of $E\dr{s}$ and $t$ within reasonable
limits.

As it can be inferred from the nearly vertical lines in the
$\eta_1$--$\eta_2$ plane, the surface energy coefficient and thickness
depend mostly on $\eta_1$ and are rather independent of $\eta_2$. The
consequence of increasing $\eta_1$ is a reduction of the values of
$E\dr{s}$ and $t$. The lines of constant $t$ turn out to be, roughly
speaking, parallel to the lines of constant $E\dr{s}$. This means that
from the interplay of the parameters $\eta_1$ and $\eta_2$ it is not
possible to change the value of $t$ relative to that of $E\dr{s}$ (for
example, we see in Figure 8 that $t \sim 2.2$ fm if $E\dr{s} = 18$
MeV). We have calculated the spin--orbit potential $V\dr{so}(z)$ at
the Fermi surface for several values of $\eta_1$ and $\eta_2$. We have
found that these couplings have a marginal effect on the spin--orbit
strength, as it happened to be the case with the other bulk
non-linearity $\eta_0$.

To get some information about the incidence on the surface energy and
thickness of all the volume non-linear meson interactions together, we
have repeated the calculations in the $\eta_1$--$\eta_2$ plane setting
$\eta_0= 2$ for the quartic vector self-interaction. One finds similar
features to those of Figure 8. The effect of $\eta_0= 2$ is just
shifting $E\dr{s}$ and $t$ towards smaller values as compared with the
case $\eta_0= \infty$ ($\zeta_0 = 0$), which is in accordance with
what was found in Figures 2 and 3 at $K= 200$ MeV and
$M_{\infty}^{*}/M= 0.6$.

%
%
\subsection{Influence of the non-linear terms with gradients}
\hspace*{\parindent}
Now we discuss the non-linear interactions
%
\beq
\frac{1}{2} \frac{\Phi}{M}
\lf[ \frac{\alpha_1}{g\dr{s}^2}
\lf( \mbox{\boldmath $\nabla$}\Phi \ri)^2
- \frac{\alpha_2}{g\dr{v}^2}
\lf( \mbox{\boldmath $\nabla$}W \ri)^2 \ri]
\label{eq3.8} \eeq
that vanish in infinite nuclear matter. We recall that these terms are
actually of order 5 in the expansion of the effective Lagrangian but,
following Refs.\ \cite{Ser97} and \cite{Fur97}, we include them
because they can be relevant in the surface due to their gradient
structure.

Using the same saturation properties and scalar mass of Figure 8, in
Figure 9 we have calculated $E\dr{s}$ and $t$ for several values of
$\alpha_1$ and $\alpha_2$ with $\zeta_0= \eta_1= \eta_2= f\dr{v}= 0$.
One observes that the curves of constant $E\dr{s}$ are projected onto
the plane $\alpha_1$--$\alpha_2$ as almost parallel straight lines (at
least in the analyzed region, corresponding to natural values of
$\alpha_1$ and $\alpha_2$). The same happens to the curves of constant
$t$. But in contrast with the situation found in the plane
$\eta_1$--$\eta_2$ (Figure 8), the slope of the lines of constant $t$
is different from that of the lines of constant $E\dr{s}$. This means
that by varying $\alpha_1$ and $\alpha_2$ one can achieve some
modification on the surface thickness while keeping the same surface
energy. For example, if we consider the contour line of $E\dr{s}= 18$
MeV we find that for $\alpha_2= 2.0$ it is $t \sim 2.05$ fm, whereas
for $\alpha_2= -1.5$ it is $t \sim 2.25$ fm. From Figure 9 we also see
that increasing $\alpha_1$ at constant $\alpha_2$ brings about larger
values of $E\dr{s}$ and $t$, and that the opposite happens if one
increases $\alpha_2$ at constant $\alpha_1$.

We have repeated the calculations of Figure 9 ($K= 200$ MeV,
$M_{\infty}^{*}/M= 0.6$) for $K= 350$ MeV and for $M_{\infty}^{*}/M=
0.7$, to verify to which extent the behaviour in the
$\alpha_1$--$\alpha_2$ plane is affected by the incompressibility and
effective mass of the interaction. Certainly, the contour lines of
$E\dr{s}$ and $t$ are shifted with respect to Figure 9, but the trends
with $\alpha_1$ and $\alpha_2$ turn out to be qualitatively the same.
The range of variation of the surface energy and thickness in the
$\alpha_1$--$\alpha_2$ region we are considering is shorter when
$M^{*}_\infty/M= 0.7$, while it is more or less the same when $K= 350$
MeV\@.

To assess the importance of the bulk non-linear meson interactions on
our study on $\alpha_1$ and $\alpha_2$, we have performed calculations
as in Figure 9 but setting $\eta_0= 2$ with $\eta_1= \eta_2= 0$, and
setting $\eta_1= 1$ with $\eta_0= \infty$ and $\eta_2= 0$ (as
indicated, the effect of $\eta_2$ is much smaller than that of
$\eta_1$). The results show a completely similar behaviour to Figure
9. Even the slope of the contour lines of $E\dr{s}$ and $t$ in the
$\alpha_1$--$\alpha_2$ plane changes only slightly. Comparing with
Figure 9, when $\eta_0= 2$ one finds that $E\dr{s}$ is shifted by
approximately $-1$ MeV, and that when $\eta_1= 1$ then $E\dr{s}$ is
shifted by around $-3$ MeV\@. The shifts of the surface thickness $t$
are less regular and their magnitude depends on the value of
$\alpha_1$ and $\alpha_2$.

In order to investigate the impact of the gradient interactions
$\alpha_1$ and $\alpha_2$ on the spin--orbit potential, in Figure 10
we have plotted $V\dr{so}(z)$ at the Fermi surface for a few selected
values of $\alpha_1$ and $\alpha_2$. The nuclear matter properties and
the scalar mass are the same as in Figure 6, where we studied the
dependence of $V\dr{so}(z)$ on $\eta_0$. One can see that the meson
interaction with $\alpha_1= 1$ and $\alpha_2= 0$ reduces the strength
of $V\dr{so}(z)$ and shifts the position of the minimum slightly to
the exterior. On the contrary, the interaction with $\alpha_1= 0$ and
$\alpha_2= 1$ makes the potential well deeper. The combined effect is
probed in the case $\alpha_1= \alpha_2= 1$. Since in the relativistic
model the spin--orbit force is strongly correlated with the Dirac
effective mass, we compare in Figure 10 the situation at
$M^{*}_{\infty}/M = 0.6$ and at $M^{*}_{\infty}/M = 0.7$. We realize
that the incidence of $\alpha_1$ on $V\dr{so}(z)$ is weaker for
$M^{*}_{\infty}/M = 0.7$. The small perturbations arising from the
gradient interactions when $M^{*}_{\infty}/M = 0.7$ are not sufficient
to produce a spin--orbit strength equivalent to that of the case
$M^{*}_{\infty}/M = 0.6$.

%
%
\subsection{Role of the tensor coupling of the omega meson}
\hspace*{\parindent}
To conclude this section we investigate the influence of the $\omega$
tensor coupling
%
\beq
\sum_\alpha \varphi_\alpha^\dagger(z)
\lf[ - \frac{i f\dr{v}}{2M} \, \beta
  \mbox{\boldmath$\alpha$} \!\cdot\! \mbox{\boldmath$\nabla$} W(z)
\ri] \varphi_\alpha (z)
\label{eq3.9} \eeq
which adds some momentum and spin dependence to the interaction. The
natural combination for this coupling is $f\dr{v}/4$.

Well known from one-boson-exchange potentials (where $f\dr{v}$ above
is commonly written as $f\dr{v}/g\dr{v}$), the tensor coupling was
included in the fits to nuclear properties of Refs.\
\cite{Rei89,Ruf88} (conventional QHD) and \cite{Ser97,Fur97}
(effective field theory), and in the study of the nuclear spin--orbit
force in chiral effective field theories carried out in Ref.\
\cite{Fur98}. These works noticed the existence of a trade-off between
the size of the $\omega$ tensor coupling and the size of the scalar
field. In other words, the tensor coupling breaks the tight connection
existing in relativistic models between the value of the nucleon
effective mass at saturation and the empirical spin--orbit splitting
in finite nuclei (which constrains $M^{*}_{\infty}/M$ to lie between
0.58 and 0.64 \cite{Fur96}). Including a tensor coupling the authors
of Refs.\ \cite{Ser97,Fur97,Fur98} were able to obtain natural
parameter sets that provide excellent fits to nuclear properties and
spin--orbit splittings with an equilibrium effective mass remarkably
higher ($M^{*}_{\infty}/M \sim 0.7$) than in models that ignore such
coupling. We want to analyze the nature of this effect in the simpler
but more transparent framework of semi-infinite nuclear matter.

In Figure 11 we have drawn the surface energy coefficient and the
surface thickness as functions of $f\dr{v}$ in the range $[-0.6,0.9]$
for two values of the effective mass and of the incompressibility,
having set $m\dr{s}= 490$ MeV, $\rho_0 = 0.152$ fm$^{-3}$ and $a\dr{v}
= -16.42$ MeV\@. To exemplify the incidence of $f\dr{v}$ on the
spin--orbit potential, Figure 12 displays $V\dr{so}(z)$ at the Fermi
surface for a few of the cases of Figure 11. We also performed the
calculations for $m\dr{s}= 525$ MeV: $E\dr{s}$ and $t$ are shifted
downwards with respect to Figure 11 and $V\dr{so}(z)$ is deeper than
in Figure 12, but the global trends with $f\dr{v}$ are the same.

Figure 11 shows the strong reduction of $E\dr{s}$ and $t$ as $f\dr{v}$
increases (the slope of the curves is milder for $M^{*}_{\infty}/M=
0.7$ than for $M^{*}_{\infty}/M= 0.6$). Figure 12 reveals
that this fact is associated with a deeper and wider spin--orbit
potential. This agrees with the results of Hofer and Stocker
\cite{Hof89} who showed in the standard RMF model that the
spin--orbit coupling reduces the surface energy and thickness. At
variance with the individual values of $E\dr{s}$ and $t$, the ratio
$E\dr{s}/t$ stays to a certain extent constant with $f\dr{v}$.

Figure 12 evinces the sensitivity of $V\dr{so}(z)$ to $f\dr{v}$. The
lower the nucleon effective mass is, the larger the effect. For
$M^{*}_{\infty}/M= 0.7$ we realize that with positive values of
$f\dr{v}$ ($\sim 0.3$ in the present case) one can get a spin--orbit
strength comparable, or even stronger, to that of the case
$M^{*}_{\infty}/M= 0.6$ and $f\dr{v}= 0$, something that could not be
achieved with natural values of the couplings studied in the previous
sections. Since our parametrization with $M^{*}_{\infty}/M= 0.7$ and
$K= 200$ MeV at $f\dr{v}= 0$ already has reasonable surface energy and
thickness ($E\dr{s}= 16.6$ MeV and $t= 1.97$ fm), increasing $f\dr{v}$
results in smaller values of $E\dr{s}$ and $t$. This should be
compensated with the other couplings (especially $\alpha_1$ and
$\alpha_2$) that modify the spin--orbit strength to a lesser degree
than $f\dr{v}$, or the starting point should have other values of the
incompressibility $K$ and the scalar mass $m\dr{s}$.

The spin--orbit effect has to do with the explicit dependence of the
nucleon orbital wave functions on the spin orientation $\lambda$. As
described in Ref.\ \cite{Hof89} nucleons with $\lambda= +1$ feel an
attractive spin--orbit potential and are pushed to the exterior of the
surface, whereas the spin--orbit force is repulsive for nucleons with
$\lambda= -1$ which are pushed to the interior. As a consequence of
this a depletion of particles with $\lambda= -1$ occurs at the
surface. This behaviour is contrasted in Figure 13 for $f\dr{v}= 0$
and $f\dr{v}= 0.6$ in the case $M^{*}_{\infty}/M= 0.7$. The figure
depicts the profiles of the total baryon and tensor densities as well
as those of their spin components $\rho^\lambda (z)$ and
$\rho\dr{T}^\lambda (z)$ for $\lambda= \pm1$, Eqs.\
\req{eq7}--\req{eq9b}. When the spin--orbit strength is large,
attraction dominates over repulsion and more particles accumulate at
the surface than particles are removed from it. Then the total baryon
density is enhanced at the surface region and it falls down more
steeply.

\pagebreak
%
\section{Surface properties in the asymmetric case}
\hspace*{\parindent}
We briefly recall some basic definitions concerning nuclear
surface symmetry properties (further details on the relativistic
treatment of asymmetric infinite and semi-infinite nuclear matter can
be found in Refs.\ \cite{Von94a,Cen98,Von95}). For a bulk neutron
excess $\delta_0= (\rho\dr{n0} - \rho\dr{p0}) / (\rho\dr{n0} +
\rho\dr{p0})$ (i.e., the asymptotic asymmetry far from the surface), a
surface energy coefficient can be computed as
%
\beq
 E\dr{s} (\delta_0) = 4\pi r_{0}^2 \int^{\infty}_{-\infty} dz
 \lf[ {\cal E}(z) - \lf( a\dr{v}(\delta_0)+M \ri) \rho(z) \ri] ,
\label{eq4.3}\eeq
where ${\cal E}(z)$ is the total energy density of the system of
neutrons and protons, $a\dr{v}(\delta_0)$ denotes the energy per
particle in nuclear matter of asymmetry $\delta_0$, and $\rho(z)=
\rho\dr{n}(z) + \rho\dr{p}(z)$ with $\rho\dr{n}$ and $\rho\dr{p}$
referring to the neutron and proton densities, respectively. According
to the liquid droplet model (LDM) \cite{Mye69}, for small values of
the neutron excess $E\dr{s} (\delta_0)$ can be expanded as follows:
%
\beq
 E\dr{s}(\delta_0) = E\dr{s} + \lf(\frac{9J^2}{4Q}
 + \frac{2E\dr{s}L}{K} \ri) \delta_0^2 + \cdots \,.
\label{eq4.4}\eeq
In this equation $J$ stands for the bulk symmetry energy coefficient,
$L$ reads for the LDM coefficient that expresses the density
dependence of the symmetry energy, and $Q$ is the so-called surface
stiffness coefficient that measures the resistance of the system
against pulling the neutron and proton surfaces apart. All of these
macroscopic coefficients are familiar from semi-empirical LDM mass
formulae.

Another quantity of interest is the neutron skin thickness $\Theta$,
namely the separation between the neutron and proton surface
locations:
%
\beq
 \Theta = \int_{-\infty}^\infty dz
 \lf[ \rho\dr{n}(z)/\rho\dr{n0} - \rho\dr{p}(z)/\rho\dr{p0} \ri] .
\label{eq4.5}\eeq
In finite nuclei $\Theta$ would correspond to the difference between
the equivalent sharp radii of the neutron and proton distributions. In
the small asymmetry limit the LDM predicts a linear behaviour of
$\Theta$ with $\delta_0$:
%
\beq
 \Theta = \frac{3 r_0}{2} \frac{J}{Q} \delta_0 \,.
\label{eq4.6}\eeq

For calculations of finite nuclei of small overall asymmetry $I =
(N-Z)/A$, the LDM expansion of the energy can be written as
%
\beq
 E = \lf( a\dr{v} + J I^2 \ri) A
 + \lf[ E\dr{s} - \lf( \frac{9J^2}{4Q}
        - \frac{2E\dr{s}L}{K} \ri) I^2 \ri] A^{2/3}
 + a\dr{C} Z^2 A^{-1/3} + \cdots \,,
\label{eq4.7}\eeq
where $a\dr{C}$ is the Coulomb energy coefficient. Notice that $I \neq
\delta_0$ in finite nuclei.

To describe asymmetric matter in the relativistic approach we need to
generalize the energy density \req{eq1} by including the isovector
$\rho$ meson. In terms of the mean field $R= g_\rho b_0$, with $b_0$
the time-like neutral component of the $\rho$-meson field, the
additional contributions to Eq.\ \req{eq1} read
%
\beqa
 & &
  \sum_\alpha \varphi_\alpha^\dagger(z) \lf[ - \frac{i f_\rho}{4M}
  \, \tau_3 \, \beta \mbox{\boldmath$\alpha$} \!\cdot\!
  \mbox{\boldmath$\nabla$} R(z) \ri] \varphi_\alpha (z)
+ \frac{1}{2} R(z) \lf[ \rho\dr{p}(z) - \rho\dr{n}(z) \ri]
\nonumber \\[3mm]
 & & \null
- \frac{1}{2g_\rho^2} \lf( \mbox{\boldmath $\nabla$} R(z) \ri)^2
- \frac{1}{2} \lf( 1 + \eta_\rho \frac{\Phi(z)}{M} \ri)
  \frac{m_\rho^2}{g_\rho^2} R^2(z) \,.
\label{eq4.1}\eeqa
The symmetry energy coefficient turns out to be
%
\beq
 J = \frac{k\dr{F}^2}{ 6 \, (k\dr{F}^2 + {M^*_\infty}^2 )^{1/2} }
    + \frac{g_\rho^2 k\dr{F}^3}{12\pi^2 m_\rho^2} \,
     \frac{1}{ 1 + \eta_\rho (1 - M^*_\infty/M) } \,.
\label{eq4.2}\eeq
In the conventional model one has $f_\rho= \eta_\rho= 0$. The
isovector tensor coupling $f_\rho$ was included in the calculations of
Refs.\ \cite{Rei89,Ser97,Fur97,Ruf88}. The new non-linear coupling
$\eta_\rho$ between the $\rho$- and $\sigma$-meson fields is of order
3 in the expansion and it has been introduced in Refs.\
\cite{Ser97,Fur97}. We will not consider higher-order non-linear
couplings involving the $\rho$ meson since the expectation value of
the $\rho$ field is typically an order of magnitude smaller than that
of the $\omega$ field \cite{Ser97,Fur97}. For example, in calculations
of the high-density nuclear equation of state, M\"uller and Serot
\cite{Mul96} found the effects of a quartic $\rho$ meson coupling
($R^4$) to be only appreciable in stars made of pure neutron matter.
On the other hand, in analogy to the couplings $\alpha_1$ and
$\alpha_2$ for the $\sigma$ and $\omega$ fields, we also tested a
surface contribution $-\alpha_3 \Phi \, ( \mbox{\boldmath$\nabla$} R
)^2 /(2 g_\rho^2 M)$ and found that the impact it has on the
properties we will study in this section is absolutely negligible.

As we have seen, the quantity that governs the surface properties in
the regime of low asymmetries is the surface stiffness $Q$. Table 1
analyzes the effect on $Q$ and $L$ of the couplings discussed in the
preceding sections and of the $f_\rho$ and $\eta_\rho$ parameters. On
the basis of Eq.\ \req{eq4.6}, we have extracted $Q$ from a linear
regression in $\delta_0$ to fit our results for $\Theta$ up to
$\delta_0= 0.1$. We have set the equilibrium properties to $\rho_0=
0.152$ fm$^{-3}$, $a\dr{v}= -16.42$ MeV, $K= 200$ MeV,
$M^{*}_{\infty}/M= 0.6$ and $J= 30$ MeV, and have used a scalar mass
$m\dr{s}= 490$ MeV and a $\rho$-meson mass $m_\rho= 763$ MeV\@. Though
here we are interested in tendencies rather than in absolute values,
for comparison we mention that NL1 has (units in MeV) $J= 43.5$, $L=
140$ and $Q= 27$ \cite{Von95}, the sophisticated droplet-model mass
formula FRDM \cite{Mol93} implies $J= 33$, $L= 0$ and $Q= 29$, and the
ETFSI-1 mass formula \cite{Abo92} based on microscopic forces predicts
$J= 27$, $L= -9$ and $Q= 112$.

Table 1 shows that the influence on the surface stiffness of the
volume self-interactions $\eta_0$, $\eta_1$ and $\eta_2$ is not very
large for natural values of these couplings. In the present case $Q$
is slightly increased by decreasing $\eta_0$ (i.e., by increasing the
quartic vector coupling $\zeta_0$). For $\eta_1= 1$ we find a
non-negligible increase of $Q$, which signals a larger rigidity of the
nuclear system against the separation of the neutron and proton
surfaces. The effect of $\eta_2$ is again moderate as compared to that
of $\eta_1$. $Q$ is augmented by a positive $\eta_\rho$ coupling,
while a negative $\eta_\rho$ induces a lower value of $Q$. Some
visible changes in $Q$ take place when the $\alpha_1$ and $\alpha_2$
gradient interactions are taken into account. Due to the opposite
behaviour of $Q$ with $\alpha_1$ and $\alpha_2$, the tendencies
compensate in a case like $\alpha_1= \alpha_2= 1$, but the net effect
is reinforced e.g.\ if $\alpha_1= -\alpha_2= 1$.

As one could expect the isoscalar tensor coupling $f\dr{v}$ has a
notable effect on $Q$, even for the relatively small value
$f\dr{v}=0.3$ that we have used in Table 1. On the contrary, $Q$ is
virtually insensitive to the isovector tensor coupling $f_\rho$. The
reason is that the derivative of the $R(z)$ field is much smaller than
that of the $W(z)$ field. In the least-square fits to ground-state
properties of Refs.\ \cite{Rei89,Ruf88} nothing was gained by the
$\rho$ tensor coupling. In any case, the best fits of Refs.\
\cite{Ser97,Fur97} have $f_\rho \approx 4$.

From Table 1 we recognize that the main changes in the coefficient $L$
arise from the $\eta_\rho$ coupling. As a rule of thumb, increasing
values of $Q$ are associated with decreasing values of $L$ for the
bulk couplings $\eta_0$, $\eta_1$, $\eta_2$ and $\eta_\rho$. Since $L$
is a bulk quantity, it is not modified by the surface interactions.

Figures 14 and 15 illustrate the dependence on asymmetry of the
neutron skin thickness, surface energy and surface thickness for some
of the cases considered in Table 1. The figures extend up to
$\delta\dr{0}= 0.3$, which widely covers the range relevant for
laboratory nuclei ($\delta_0 \leq 0.2$). As the system becomes neutron
rich we can appreciate how a neutron skin develops and $\Theta$ grows
from its vanishing value at $\delta_0= 0$. For small asymmetries the
growth is linear in $\delta_0$, as predicted by the LDM. The surface
energy coefficient grows quadratically with increasing neutron excess
and the LDM equation \req{eq4.4} is clearly a good approximation. In
general, the interactions having thicker neutron skins (smaller values
of $Q$) also have larger surface energies.

The parameter $\eta_\rho$ can be used for the fine tuning of the
symmetry properties of the interaction without spoiling the
predictions for symmetric systems. If in the conventional ansatz
$g_\rho$ is fixed by the value of the symmetry energy $J$, in the
extended model $J$ depends on a combination of $g_\rho$ and
$\eta_\rho$, Eq.\ \req{eq4.2}. Therefore, $\eta_\rho$ provides in
practice a mechanism that can help to simultaneously adjust $Q$ (to
get the required neutron skin $\Theta$) and $J$ (to keep the fit to
the masses) preserving the symmetric surface properties.

\pagebreak
%
\section{Summary}
\hspace*{\parindent}
Within relativistic mean field theory, we have investigated the
influence on nuclear surface properties of the non-linear meson
interactions and tensor couplings recently considered in the
literature. These interactions, beyond standard QHD, are based on
effective field theories. The effective field theory approach allows
one to expand the non-renormalizable couplings, which are consistent
with the underlying QCD symmetries, using naive dimensional analysis
and the naturalness assumption \cite{Ser97,Fur96,Mul96,Fur97}.

The quartic vector self-interaction $\zeta_0$ makes it possible to
obtain a desirable positive value of the coupling constant $\kappa_4$
of the quartic scalar self-interaction, for realistic nuclear matter
properties and within the bounds of naturalness. This $\zeta_0$
coupling has only a slight impact on the surface properties.
Nevertheless, it helps to find parametrizations where both the surface
energy coefficient $E\dr{s}$ ant the surface thickness $t$ lie in the
empirical region. The $\zeta_0$ vector non-linearity makes the
spin--orbit potential well deeper, although the effect is almost
negligible. Concerning the volume non-linear couplings $\eta_1$ and
$\eta_2$, they also allow one to obtain positive values of $\kappa_4$
in the region of naturalness, depending somewhat on the saturation
properties (incompressibility and effective mass). The surface
properties are not much affected by these bulk terms either, and it
turns out that $\eta_2$ has a marginal effect as compared to that of
$\eta_1$.

The equilibrium properties do not depend on the couplings $\alpha_1$
and $\alpha_2$ that involve the gradients of the fields. Thus, these
couplings serve to improve the quality of the surface properties
without changing the bulk matter. In the conventional \sw model the
only parameter not fixed by the saturation conditions is the mass of
the scalar meson. In the $\alpha_1$--$\alpha_2$ plane the lines of
constant $E\dr{s}$ have a different slope than those of constant $t$.
It is then possible to keep a fixed value of $E\dr{s}$ and to modify
the value of $t$ by choosing $\alpha_1$ and $\alpha_2$ appropriately.
The range of variation of $E\dr{s}$ and $t$ with $\alpha_1$ and
$\alpha_2$ is wider than with the volume couplings. This justifies
including these gradient terms in the energy functional in spite of
being of order 5 in the expansion. The $\alpha_1$ and $\alpha_2$
surface meson interactions also influence the spin--orbit potential,
but the effect is not extremely significant.

The effective model is augmented with a tensor coupling of the
$\omega$ meson to the nucleon. An outstanding feature is the drastic
consequences it has for the spin--orbit force. We have emphasized how
inclusion of $f\dr{v}$ permits to obtain a spin--orbit strentgh
similar to that of $M^{*}_{\infty}/M \sim 0.6$ with larger values of
the equilibrium nucleon effective mass, contrary to the phenomenology
known from models without such a coupling.

We have discussed the implications of the extra couplings of the
extended model on various surface symmetry properties. We have
restricted ourselves to the regime of low asymmetries, where the
liquid droplet model can be applied and the surface stiffness
coefficient $Q$ is the key quantity. In particular we have pointed out
the role that the non-linearity $\eta_\rho$ of the isovector
$\rho$-meson field may play in the details of the symmetry properties.

%
\section*{Acknowledgements}
\hspace*{\parindent}
The authors would like to acknowledge support from the DGICYT (Spain)
under grant PB95-1249 and from the DGR (Catalonia) under grant
GR94-1022. M. Del Estal acknowledges in addition financial support
from the CIRIT (Catalonia).

\pagebreak
%

%
\pagebreak
%
\section*{Table captions}
\begin{description}
\item[Table 1.]
The surface stiffness coefficient $Q$ and the coefficient $L$ for
several values of the couplings analyzed in the text. We have set
$\rho_0= 0.152$ fm$^{-3}$, $a\dr{v}= -16.42$ MeV, $K= 200$ MeV,
$M^{*}_{\infty}/M= 0.6$, $J= 30$ MeV and $m\dr{s}= 490$ MeV\@.
\end{description}

\pagebreak
%
\section*{Table 1}
\vspace{2cm}
\begin{center}
\begin{tabular}{crrrrrrrcc}
\hline
$\eta_0$ & $\eta_1$   & $\eta_2$   & $\eta_\rho$
         & $\alpha_1$ & $\alpha_2$ & $f\dr{v}$
         & $f_\rho$   & $Q$ (MeV)  & $L$ (MeV)
\\
\hline
$\infty$ & 0 & 0 & 0 & 0 & 0 & 0 & 0 & 21   & 96 \\
    5    &   &   &   &   &   &   &   & 21.5 & 95 \\
    2    &   &   &   &   &   &   &   & 22   & 90 \\
$\infty$ & 1 & 0 & 0 & 0 & 0 & 0 & 0 & 24.5 & 91 \\
         & 0 & 1 &   &   &   &   &   & 22   & 93 \\
         & 1 & 1 &   &   &   &   &   & 25.5 & 89 \\
$\infty$ & 0 & 0 & 1 & 0 & 0 & 0 & 0 & 23   & 87 \\
         &   &   & $-1$ &   &   &   &   & 18   & 119 \\
$\infty$ & 0 & 0 & 0 & 1 & 0 & 0 & 0 & 17   & 96 \\
         &   &   &   & 0 & 1 &   &   & 25   & 96 \\
         &   &   &   & 1 & 1 &   &   & 19   & 96 \\
         &   &   &   & 1 & $-1$ &   &   & 16   & 96 \\
$\infty$ & 0 & 0 & 0 & 0 & 0 & 0.3 & 0 & 24 & 96 \\
         &   &   &   &   &   & $-0.3$ &   & 19 & 96 \\
$\infty$ & 0 & 0 & 0 & 0 & 0 & 0 & 5 & 21   & 96 \\
\hline
\end{tabular}
\end{center}

\pagebreak
%
\section*{Figure captions}
\begin{description}
\item[Figure 1.]
The couplings $g\dr{s}/4\pi$, $g\dr{v}/4\pi$, $\kappa_3$, $\kappa_4$
and $\zeta_0$ versus the parameter $\eta_0$ defined in Eq.\
\req{eq3.2.b}. The naturalness assumption requires all these couplings
to be of order unity. We have taken $\rho_0 = 0.152$ fm$^{-3}$
($k\dr{F}= 1.31$ fm$^{-1}$), $a\dr{v} = -16.42$ MeV and $m\dr{s}= 490$
MeV\@.
\item[Figure 2.]
Surface energy coefficient $E\dr{s}$ for several values of the
parameter $\eta_0$, $K$, $M^{*}_{\infty}/M$ and $m\dr{s}$, with
$\rho_0 = 0.152$ fm$^{-3}$ and $a\dr{v} = -16.42$ MeV\@.
\item[Figure 3.]
Surface thickness $t$ of the baryon density profile for several values
of the parameter $\eta_0$, $K$, $M^{*}_{\infty}/M$ and $m\dr{s}$, with
$\rho_0 = 0.152$ fm$^{-3}$ and $a\dr{v} = -16.42$ MeV\@.
\item[Figure 4.]
Surface energy coefficient $E\dr{s}$ and surface thickness $t$ for
several values of the parameter $\eta_0$, $M^{*}_{\infty}/M$ and
$m\dr{s}$, with $\rho_0 = 0.1484$ fm$^{-3}$, $a\dr{v} = -15.75$ MeV
and $K=200$ MeV (Ref.\ \cite{Bod91}).
\item[Figure 5.]
Baryon density $\rho(z)$ and surface tension density $\sigma(z)=
{\cal E}(z) - (a\dr{v}+M) \rho(z)$ of semi-infinite nuclear matter
for some values of the parameter $\eta_0$. It is $\rho_0= 0.152$
fm$^{-3}$, $a\dr{v}= -16.42$ MeV, $K= 200$ MeV and $m\dr{s}= 490$
MeV\@.
\item[Figure 6.]
Orbital part of the spin--orbit potential $V\dr{so}(z)$ at the Fermi
surface and central mean field $V\dr{c}(z)$, Eqs.\ \req{eq12} and
\req{eq14} respectively, for some values of the parameter $\eta_0$. It
is $\rho_0= 0.152$ fm$^{-3}$, $a\dr{v}= -16.42$ MeV, $K= 200$ MeV and
$m\dr{s}= 490$ MeV\@.
\item[Figure 7.]
The couplings $g\dr{s}/4\pi$, $g\dr{v}/4\pi$, $\kappa_3$ and
$\kappa_4$ against the parameters $\eta_1$ (left) and $\eta_2$
(right). With $\rho_0 = 0.152$ fm$^{-3}$, $a\dr{v} = -16.42$ MeV and
$m\dr{s}= 490$ MeV, results are shown for $K= 200$ MeV and
$M^{*}_{\infty}/M= 0.6$ in part (a), and for $K= 350$ MeV and
$M^{*}_{\infty}/M= 0.7$ in part (b).
\item[Figure 8.]
Level curves in the plane $\eta_1$--$\eta_2$ of the surface energy
coefficient $E\dr{s}$ (in MeV, solid lines) and of the surface
thickness $t$ (in fm, dashed lines), with $\zeta_0= \alpha_1=
\alpha_2= f\dr{v}= 0$. The point $\eta_1= \eta_2= 0$ is marked by a
cross. It is $\rho_0 = 0.152$ fm$^{-3}$, $a\dr{v} = -16.42$ MeV, $K=
200$ MeV, $M_{\infty}^{*}/M= 0.6$ and $m\dr{s}= 490$ MeV\@.
\item[Figure 9.]
Same as Figure 8 in the plane $\alpha_1$--$\alpha_2$, with $\zeta_0=
\eta_1= \eta_2= f\dr{v}= 0$. The point $\alpha_1= \alpha_2= 0$ is
marked by a cross.
\item[Figure 10.]
Orbital part of the spin--orbit potential $V\dr{so}(z)$ at the Fermi
surface for some values of the couplings $\alpha_1$ and $\alpha_2$.
The equilibrium properties of nuclear matter and the scalar mass are
the same of Figure 6.
\item[Figure 11.]
Surface energy coefficient $E\dr{s}$ and surface thickness $t$ as
functions of the strength $f\dr{v}$ of the $\omega$-meson tensor
coupling, with $\zeta_0= \eta_1= \eta_2= \alpha_1= \alpha_2= 0$. We
have set $\rho_0 = 0.152$ fm$^{-3}$, $a\dr{v} = -16.42$ MeV and
$m\dr{s}= 490$ MeV\@.
\item[Figure 12.]
Orbital part of the spin--orbit potential $V\dr{so}(z)$ at the Fermi
surface for some values of the tensor coupling $f\dr{v}$. The
equilibrium properties of nuclear matter and the scalar mass are the
same of Figures 6 and 10.
\item[Figure 13.]
Total baryon density $\rho(z)$, total tensor density $\rho\dr{T}(z)$
and their components $\rho^\lambda (z)$ and $\rho_{\rm T}^\lambda (z)$
for the two spin orientations $\lambda= \pm 1$. They have been
calculated for $f\dr{v}= 0$ and $f\dr{v}= 0.6$, with
$M^{*}_{\infty}/M= 0.7$, $K= 200$ MeV and $m\dr{s}= 490$ MeV\@.
\item[Figure 14.]
Neutron skin thickness $\Theta$ as a function of the bulk neutron
excess $\delta_0$. The solid line is the result of the conventional
model ($\zeta_0= \eta_1= \eta_2= \eta_\rho= \alpha_1= \alpha_2=
f\dr{v}= f_\rho= 0$). The other lines differ from the latter in the
indicated parameter.
\item[Figure 15.]
Same as Figure 14 for the surface energy coefficient $E\dr{s}$ and the
surface thickness $t$ as functions of the bulk neutron excess squared
$\delta_0^2$.
\end{description}
%
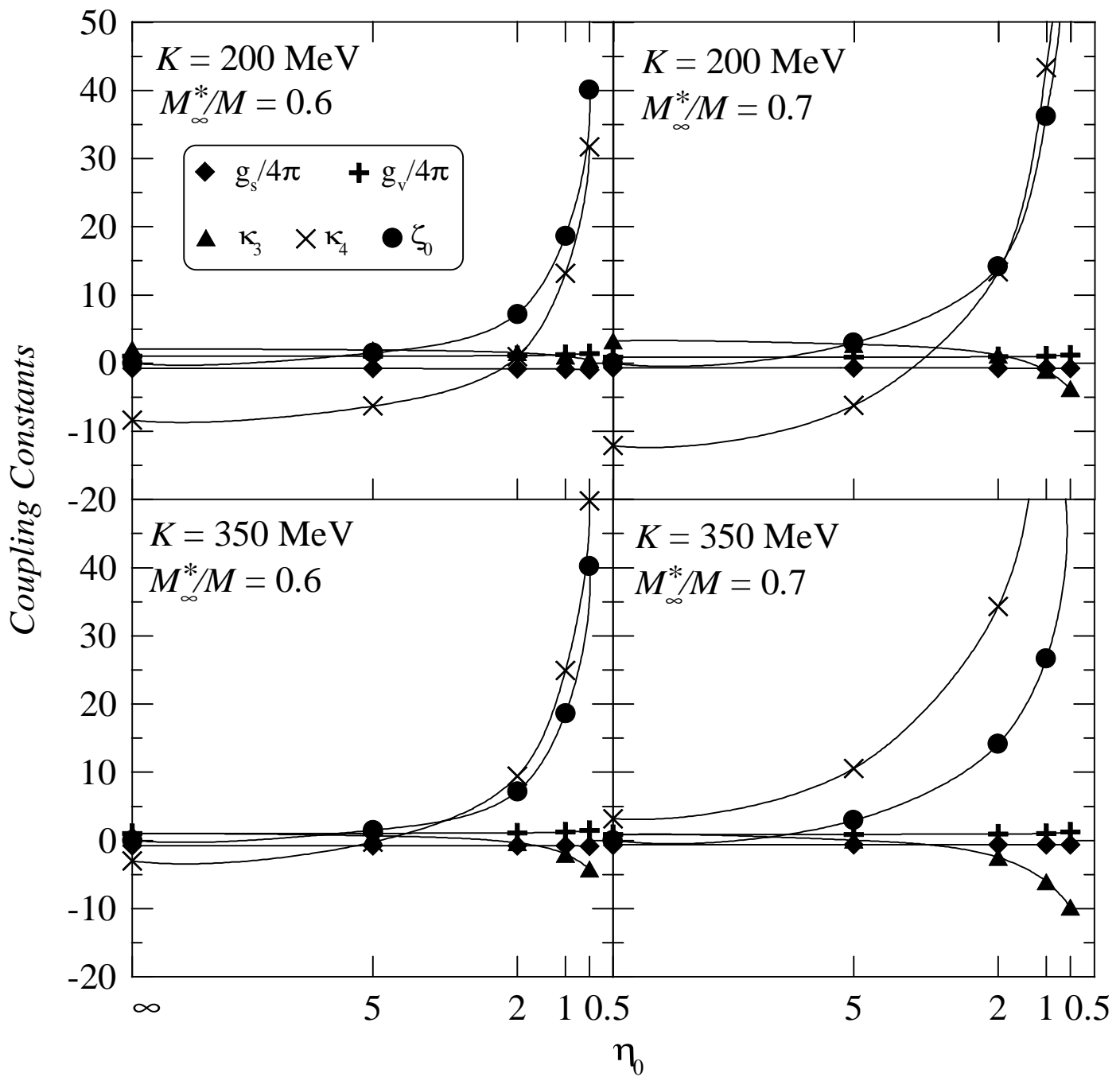
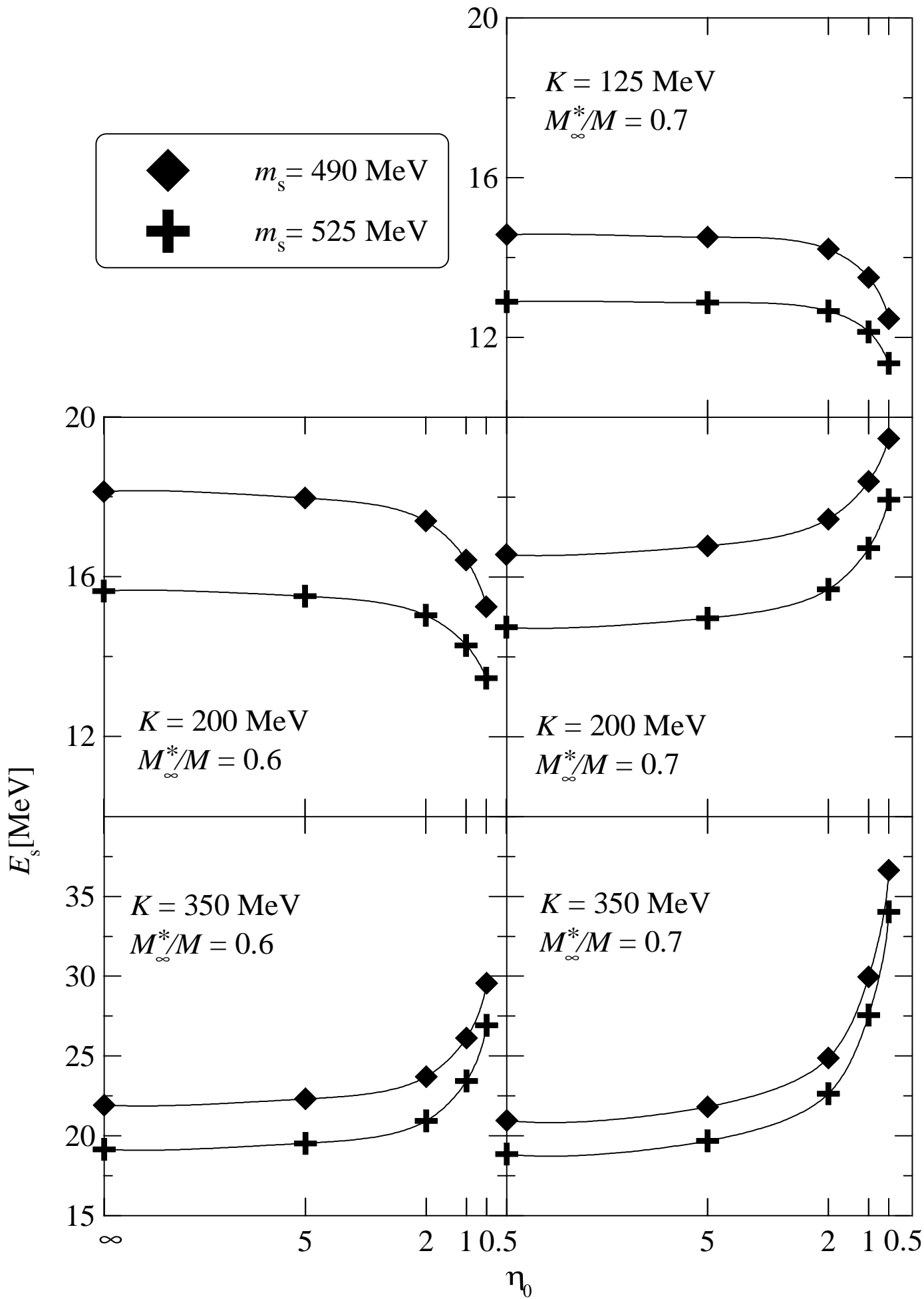
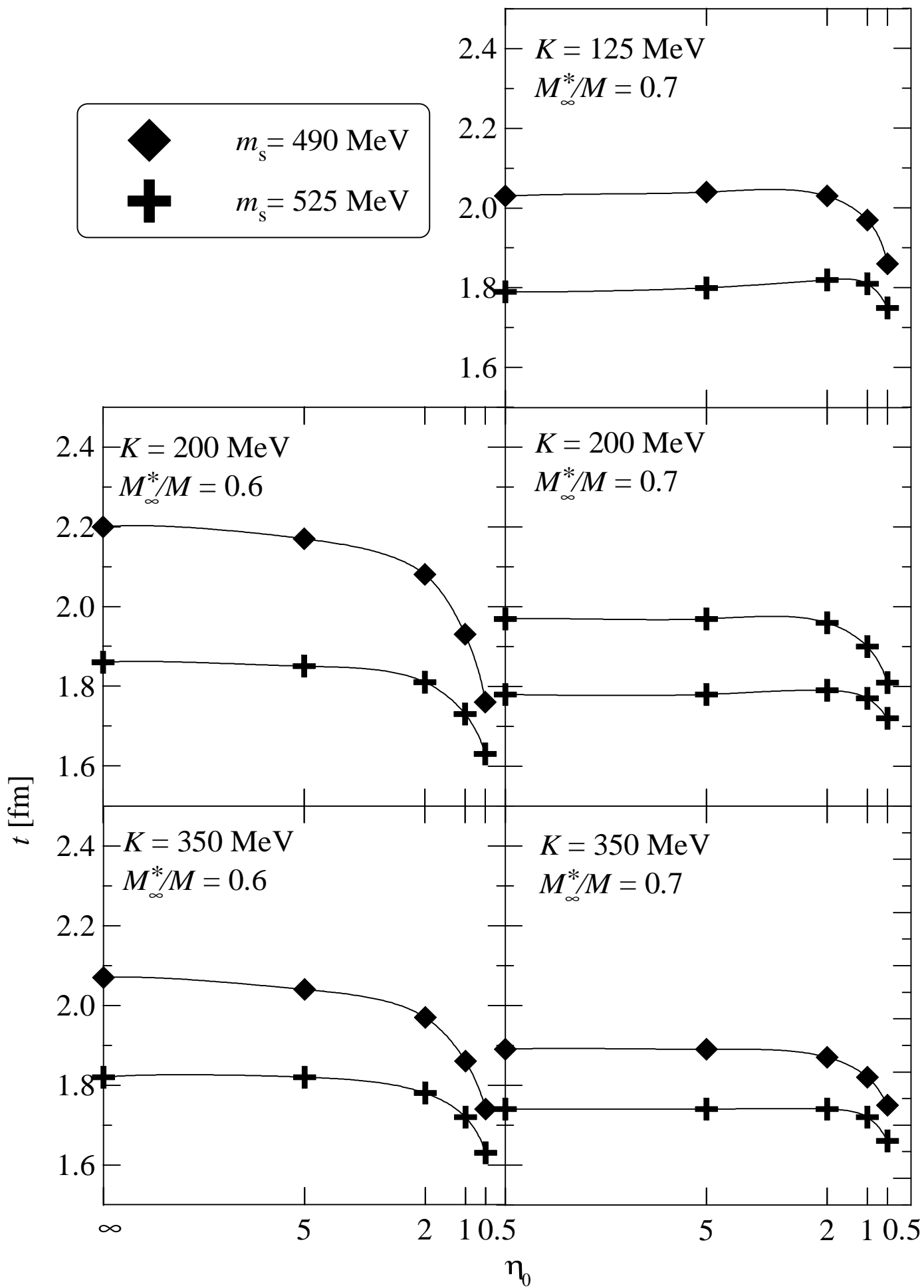
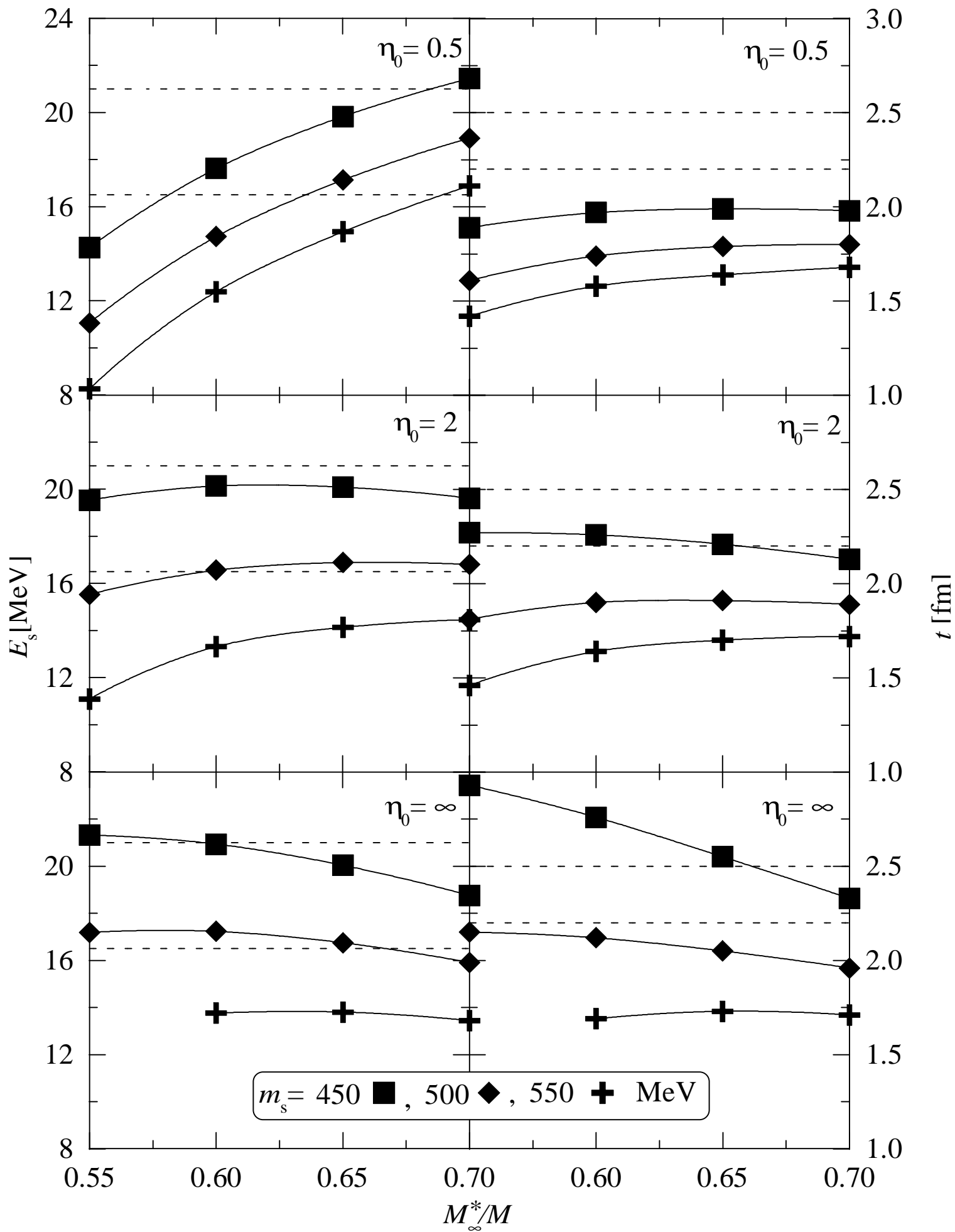
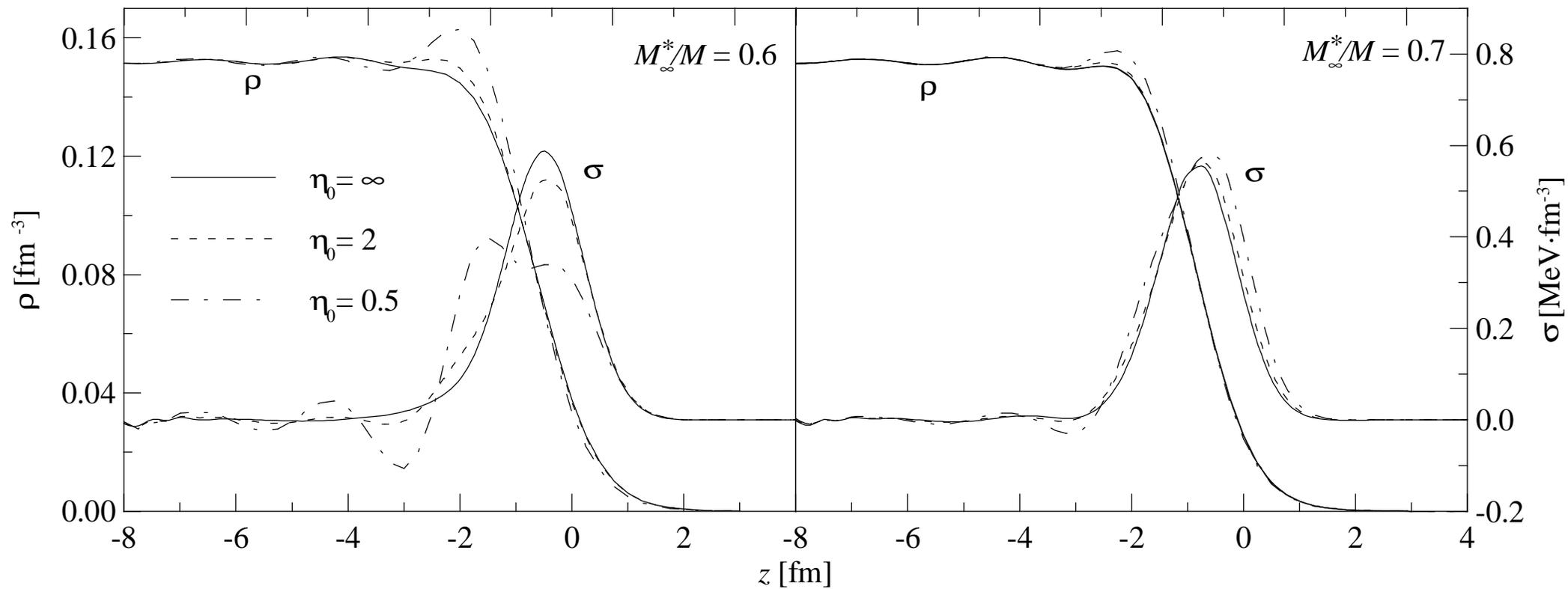
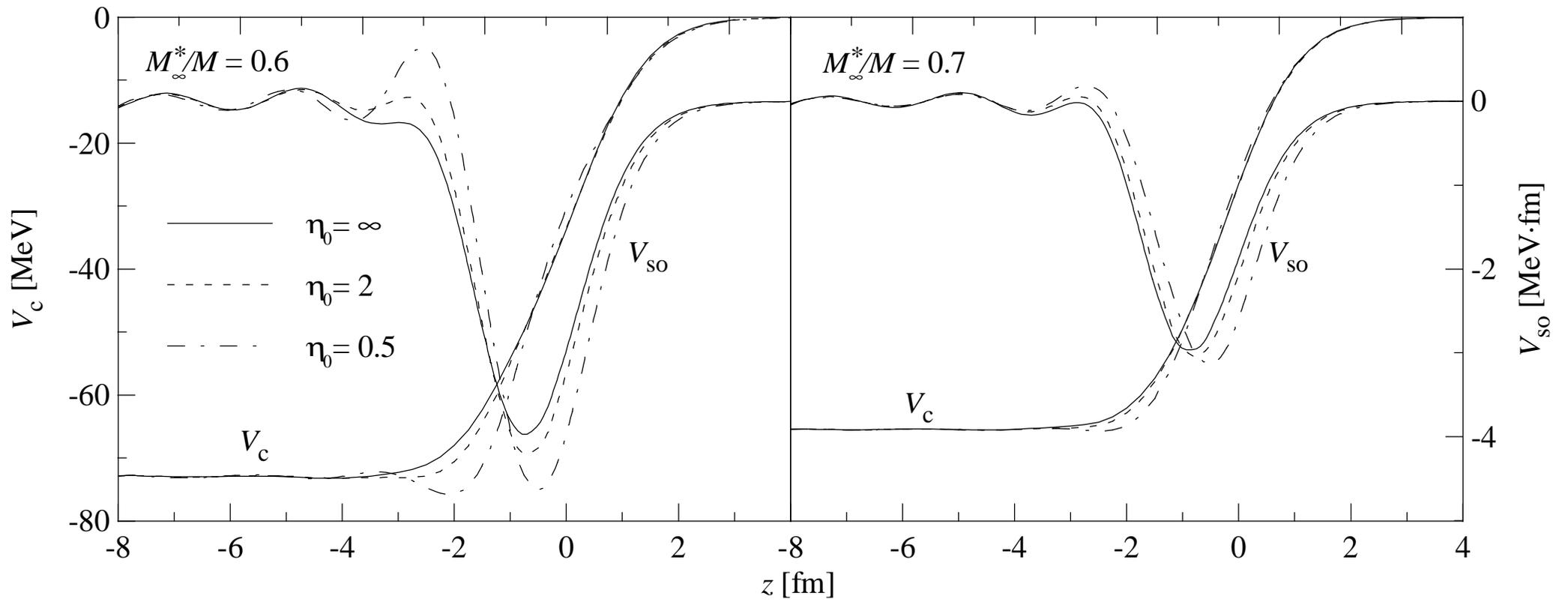
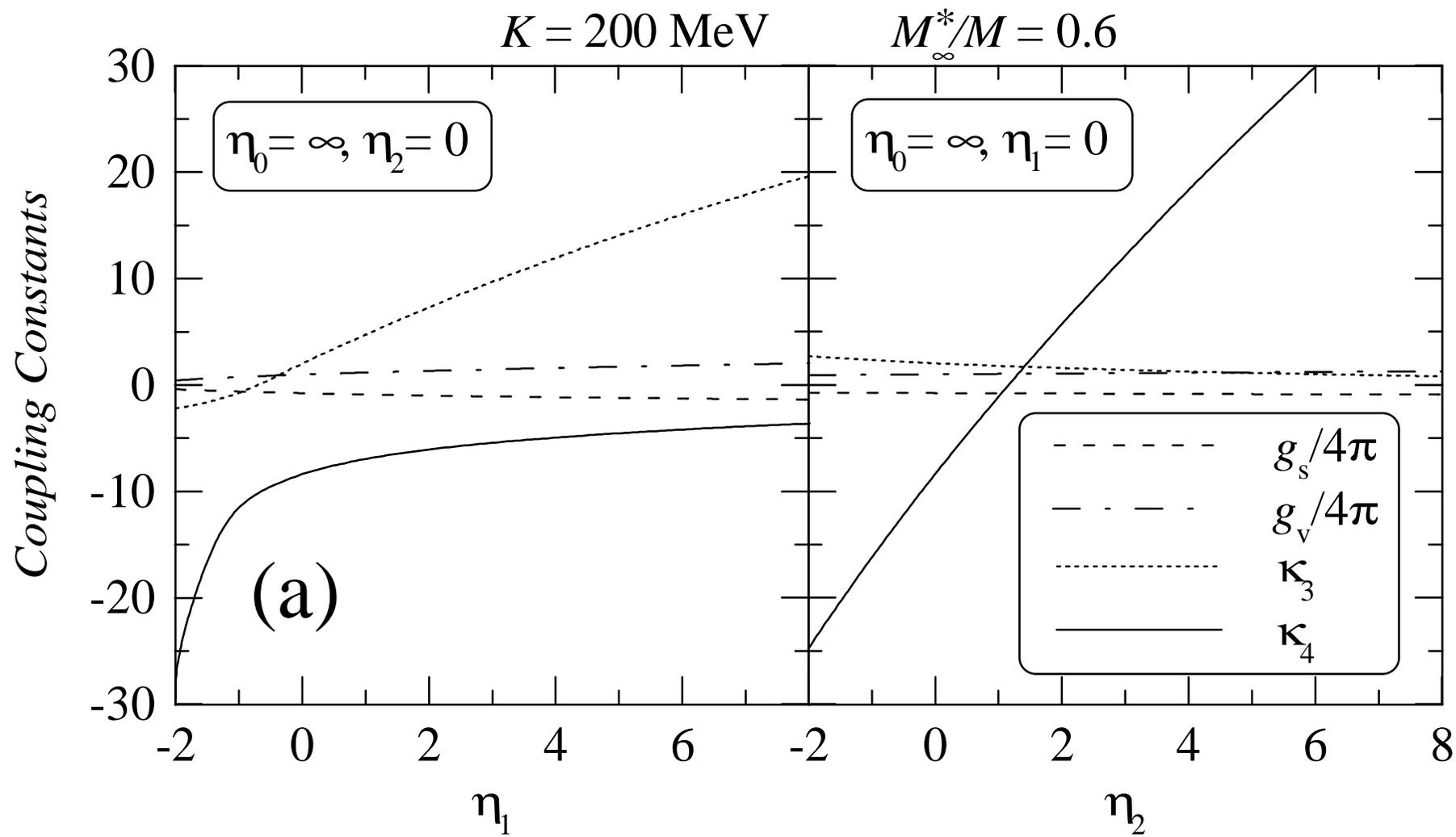
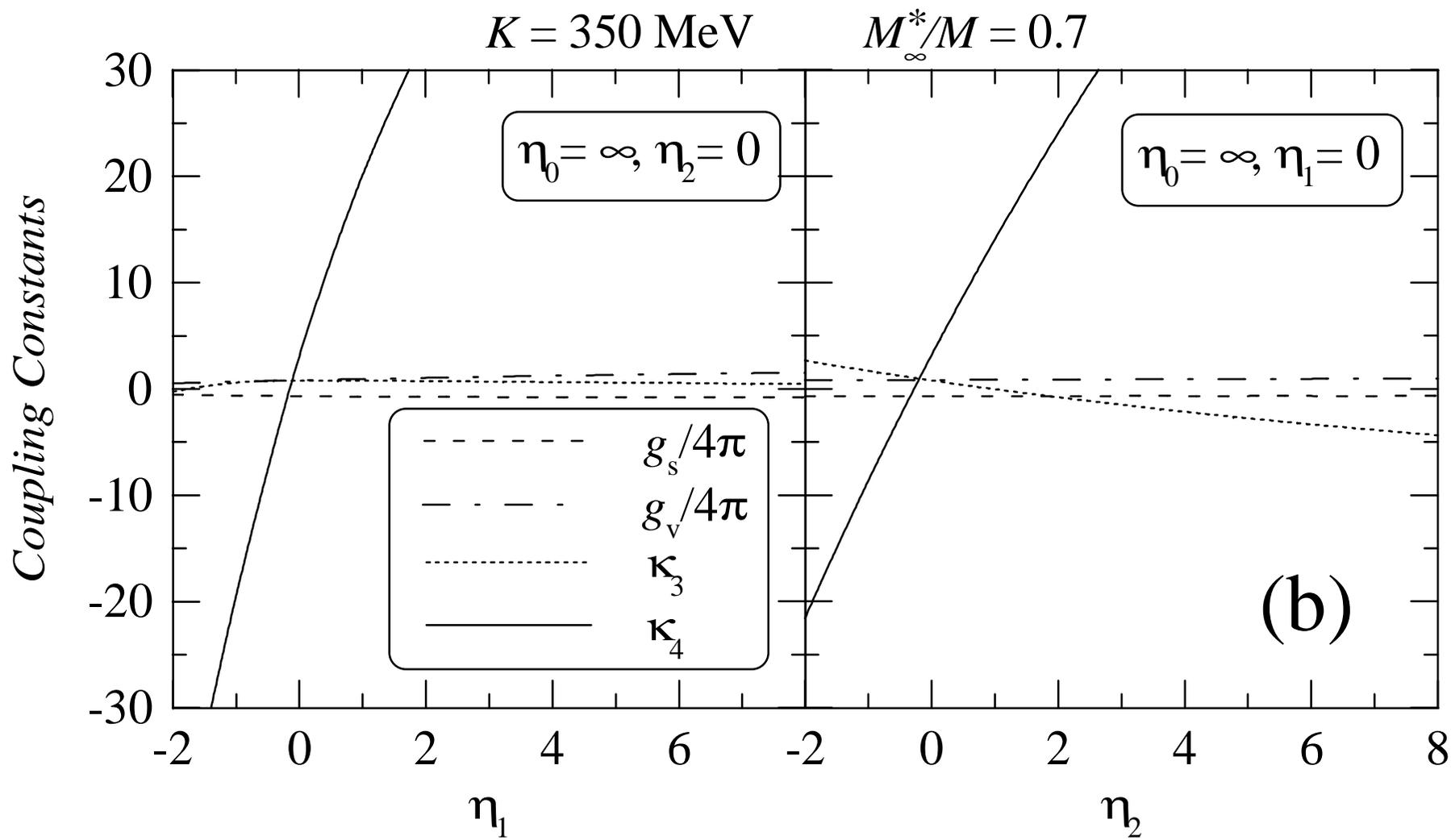
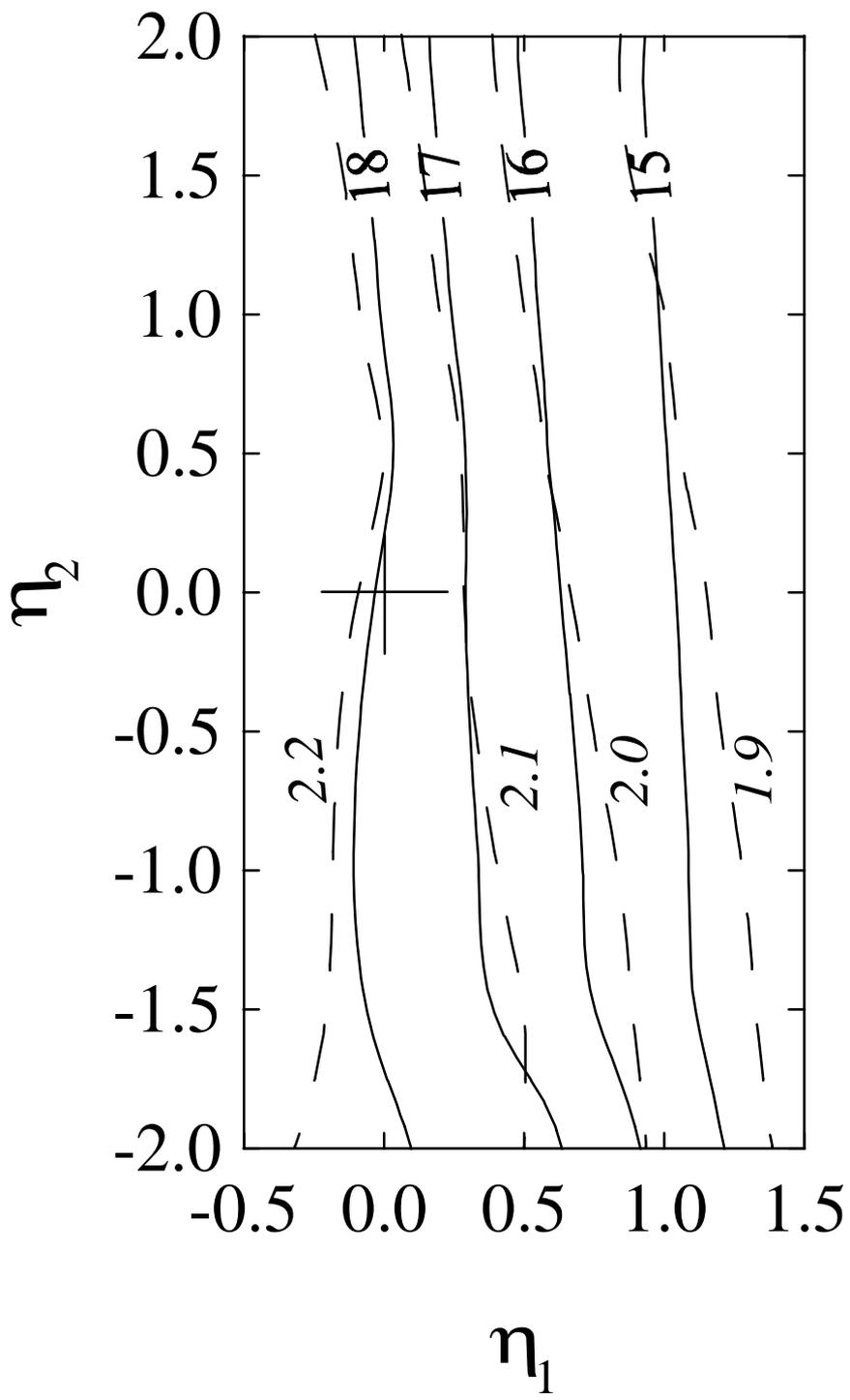
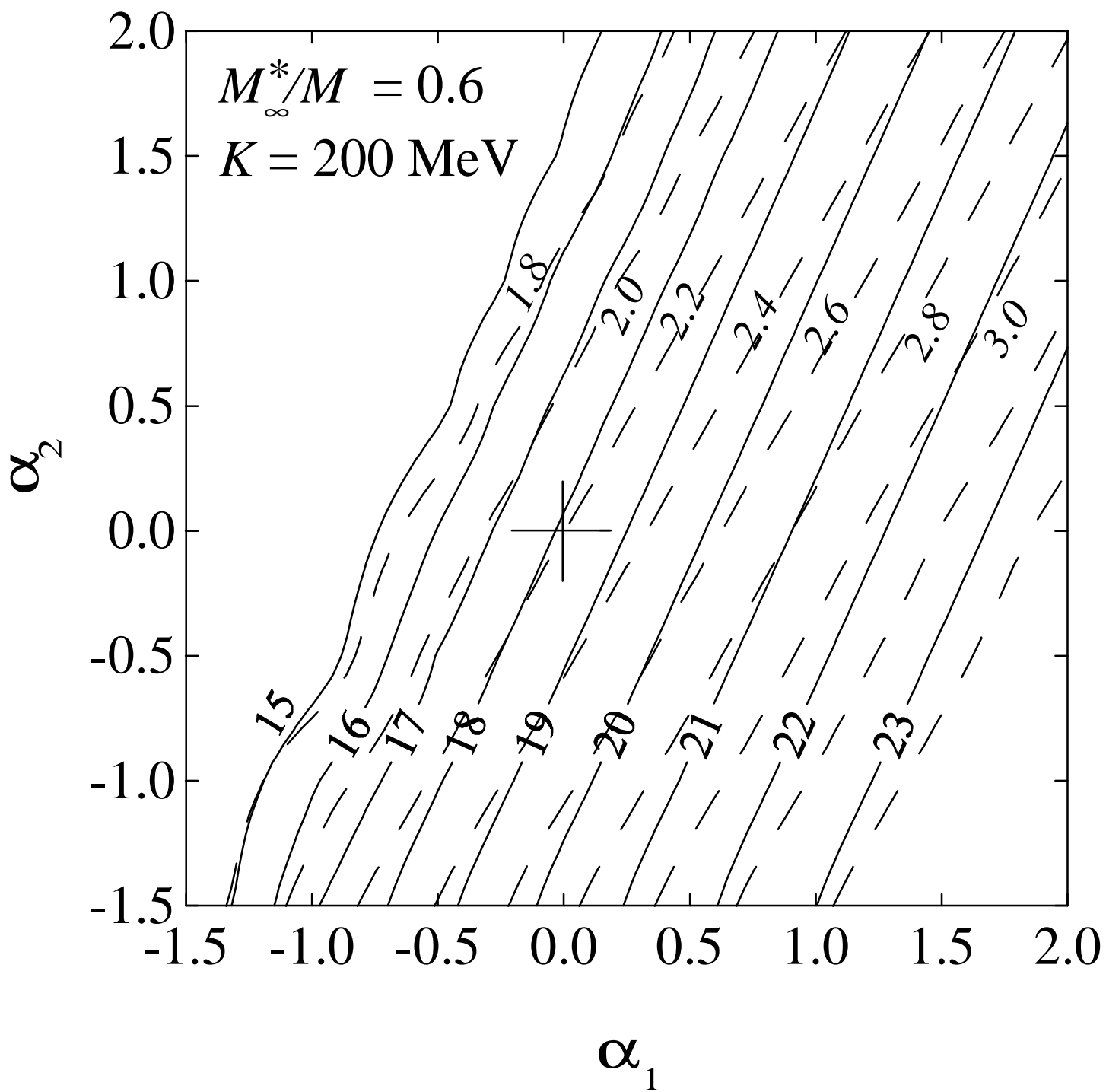
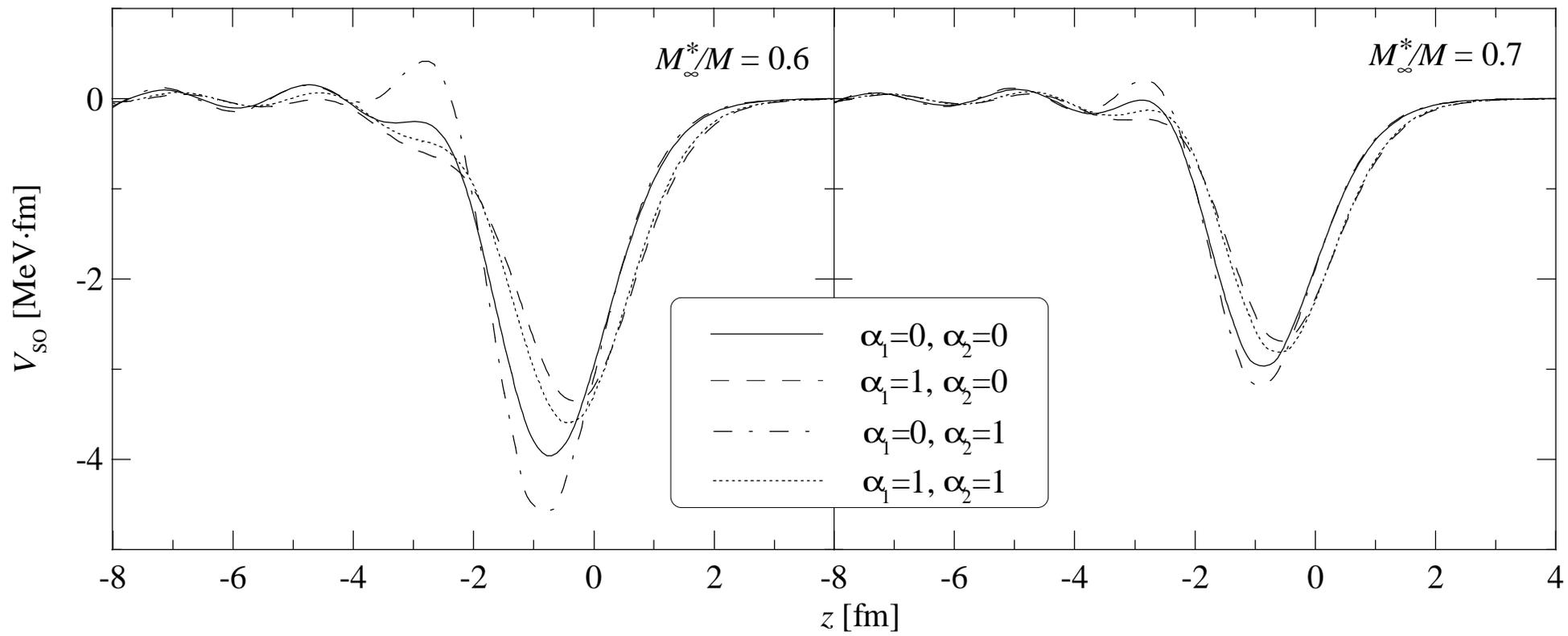
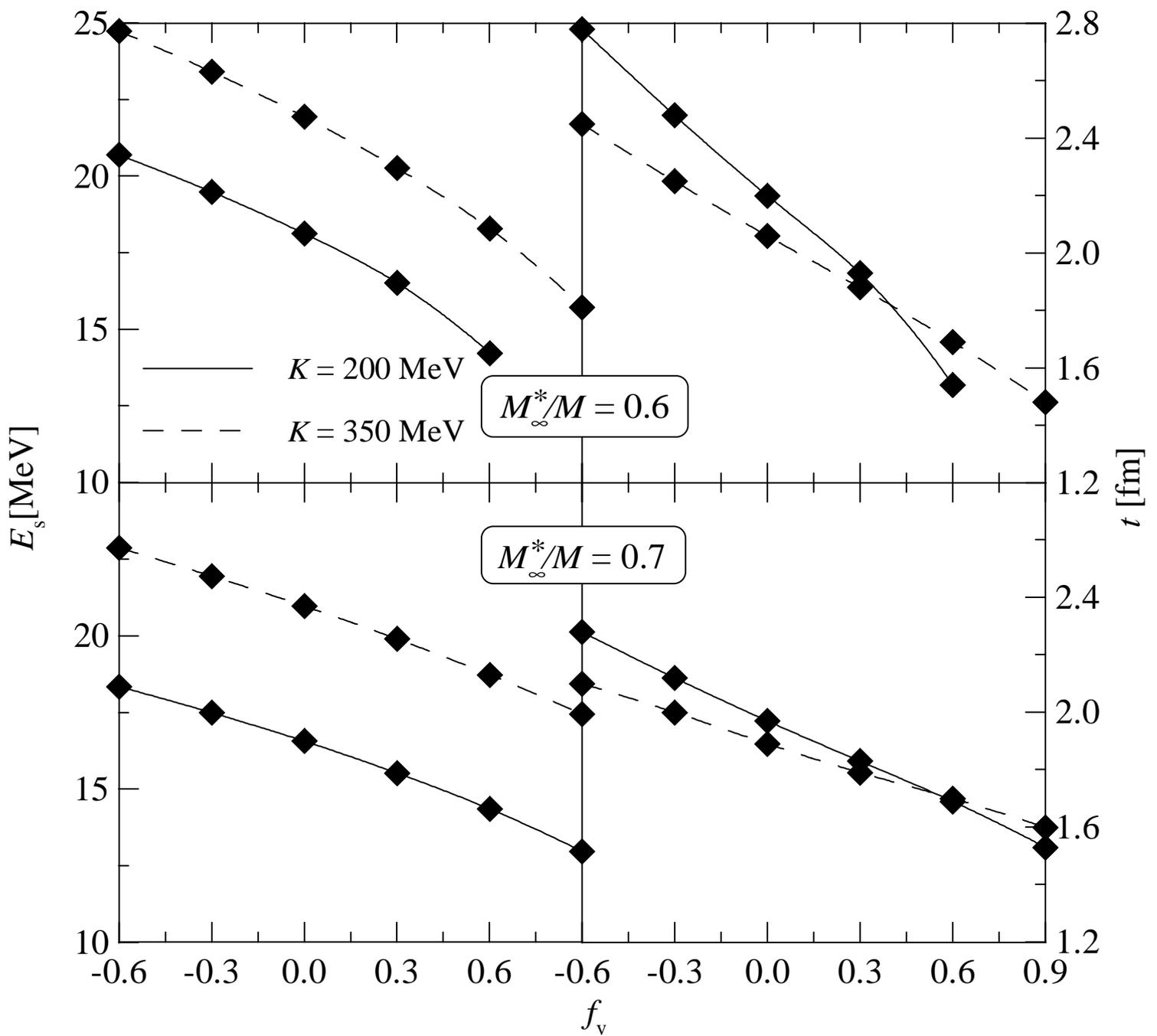
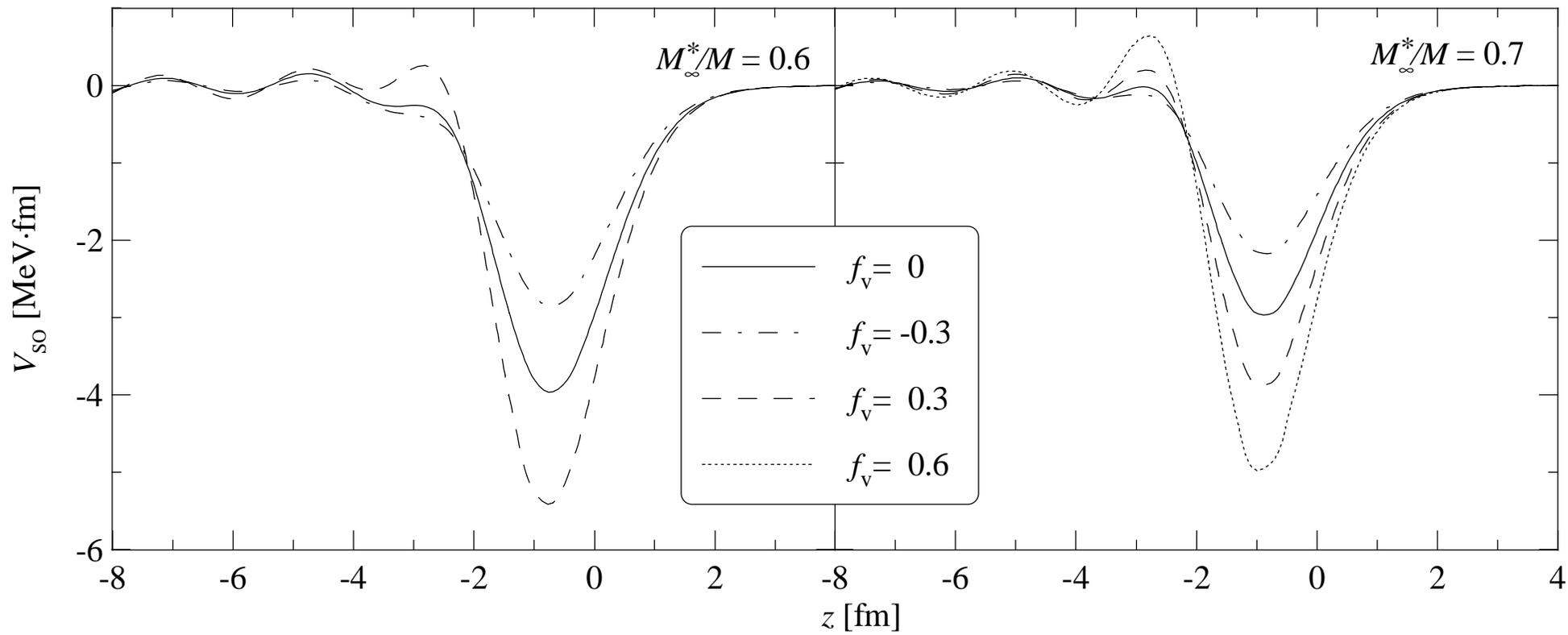
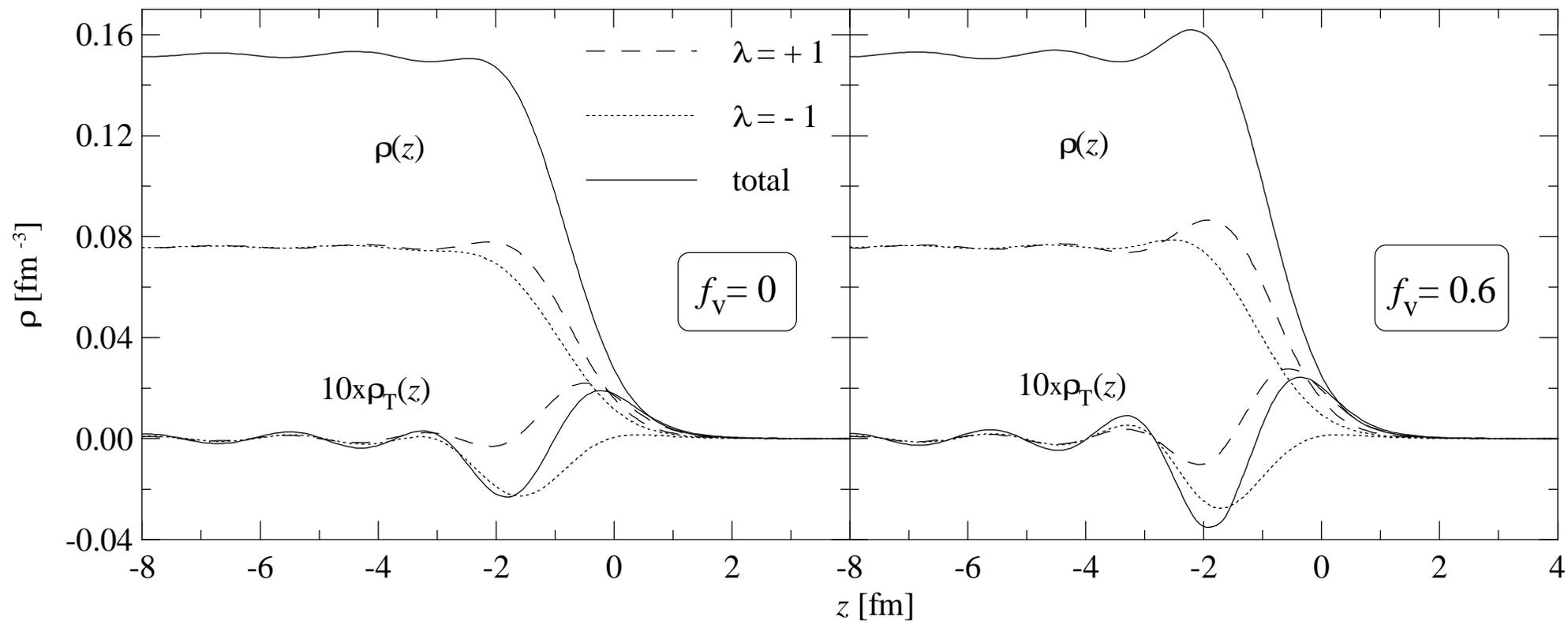
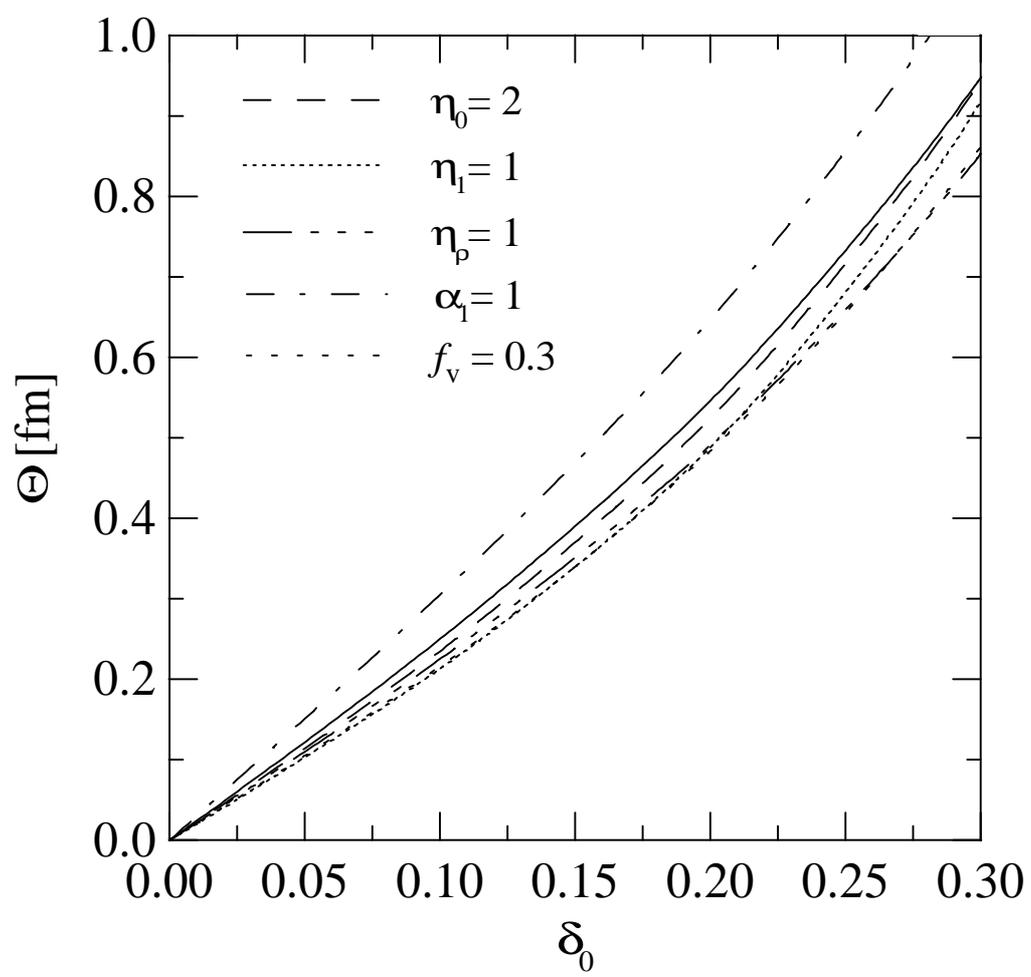
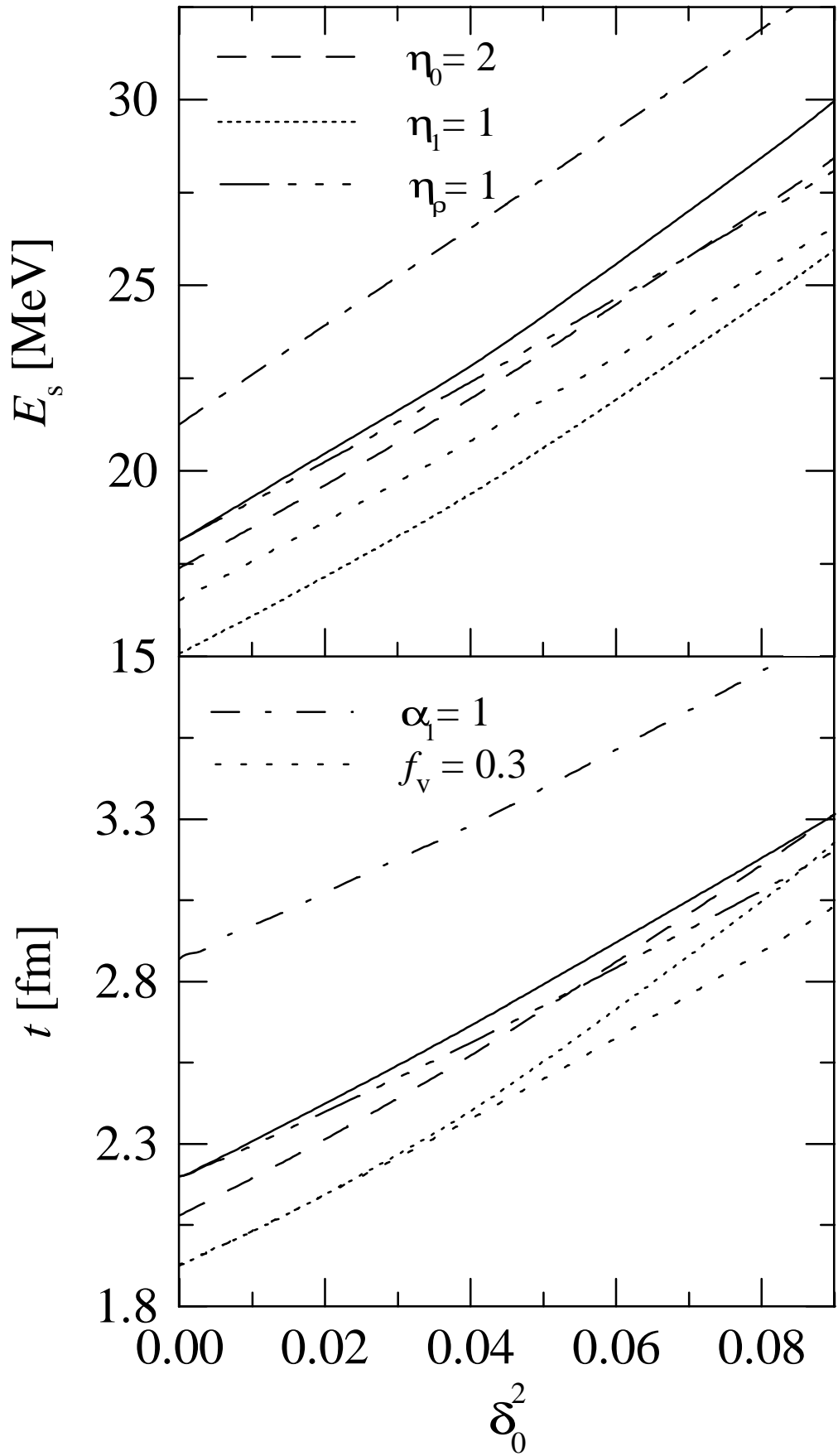
\end{document}